%% file: main.tex
\documentclass[conference]{IEEEtran}
\IEEEoverridecommandlockouts

\usepackage{cite}
\usepackage{amsmath,amssymb,amsfonts}
\usepackage{algorithmic}
\usepackage{graphicx}
\usepackage{textcomp}
\def\BibTeX{{\rm B\kern-.05em{\sc i\kern-.025em b}\kern-.08em
    T\kern-.1667em\lower.7ex\hbox{E}\kern-.125emX}}
\usepackage[hyphens]{url}
\usepackage{color,soul}
\usepackage{fancyhdr}
\usepackage[normalem]{ulem}
\usepackage[final]{microtype}
\usepackage[normalem]{ulem}
\usepackage{textcomp}
\usepackage[dvipsnames]{xcolor}
\usepackage{pifont}
\usepackage{tabularx}
\usepackage{tablefootnote}
\usepackage[normalem]{ulem}
\usepackage{array}
\usepackage{subfig}
\usepackage{enumitem}
\usepackage{setspace}
\usepackage{multirow}
\usepackage[flushleft]{threeparttable}
\usepackage{booktabs}
\usepackage{makecell}
\usepackage{tcolorbox}

\usepackage[bookmarks=true,breaklinks=true,letterpaper=true,colorlinks,citecolor=blue,linkcolor=blue,urlcolor=blue]{hyperref}
    
\newcommand{\ignore}[1]{}
\newcommand{\revision}[1]{{\color{black}{#1}}}
\newcommand{\cameraready}[1]{{\color{black}{#1}}}

\newcommand{\name}{SecDDR}
\newcommand{\papername}{\name{}}

\usepackage[pscoord]{eso-pic}% The zero point of the coordinate systemis the lower left corner of the page (the default).
\newcommand{\placetextbox}[3]{% \placetextbox{<horizontal pos>}{<vertical pos>}{<stuff>}
  \setbox0=\hbox{#3}% Put <stuff> in a box
  \AddToShipoutPictureFG*{% Add <stuff> to current page foreground
    \put(\LenToUnit{#1\linewidth},\LenToUnit{#2\paperheight}){\vtop{{\null}\makebox[0pt][c]{#3}}}%
  }%
}%

\begin{document}

%\title{Enablisng Low-Cost Secure Memories\\ by Protecting the DDR Interface}
%\title{Making Replay Attack Protection Practical \\by Securing the DDR Interface}

\title{\cameraready{\name: Enabling Low-Cost Secure Memories\\ by Protecting the DDR Interface}}
%\title{\name: Making Replay-Attack Protection Practical \\by Securing the DDR Interface}

\author{\IEEEauthorblockN{Ali Fakhrzadehgan}
\IEEEauthorblockA{\textit{The University of Texas at Austin}\\
alifakhrzadehgan@utexas.edu}\vspace*{-7mm}
\and
\IEEEauthorblockN{Prakash Ramrakhyani}
\IEEEauthorblockA{\textit{Arm}\\
prakash.ramrakhyani@arm.com}\vspace*{-7mm}
\and
\IEEEauthorblockN{Moinuddin K. Qureshi}
\IEEEauthorblockA{\textit{Georgia Tech}\\
moin@gatech.edu}\vspace*{-7mm}
\and
\IEEEauthorblockN{Mattan Erez}
\IEEEauthorblockA{
\textit{The University of Texas at Austin}\\
mattan.erez@utexas.edu}\vspace*{-7mm}}

\maketitle

\input{text/abstract}
\input{text/introduction}
\input{text/background}
\input{text/sentry}

\input{text/methodology}
\input{text/evaluation}
\input{text/compare_active_memory}
\input{text/related_work}
\input{text/conclusion}

\cameraready{
\section*{Acknowledgment}
We thank Professor Yale N. Patt and members of HPS Research Group for their inputs and providing the environment that helped shaping the early stages of this work. This work was supported in part by Intel, the Cockrell Foundation, Arm, NSF (Award \#2011145). We also thank the Texas Advanced Computing Center (TACC) for providing the computing resources.
}

\bibliographystyle{text/IEEEtranS}
\bibliography{text/refs}

\ignore{
\vspace{12pt}
\color{red}
IEEE conference templates contain guidance text for composing and formatting conference papers. Please ensure that all template text is removed from your conference paper prior to submission to the conference. Failure to remove the template text from your paper may result in your paper not being published.
}

\end{document}

%% file: text/abstract.tex
\begin{abstract}
The security goals of cloud providers and users include memory confidentiality and integrity, which requires implementing replay attack protection (RAP). RAP can be achieved using integrity trees or mutually authenticated channels. Integrity trees incur significant performance overheads and are impractical for protecting large memories. Mutually authenticated channels have been proposed only for packetized memory interfaces that address only a very small niche domain, require fundamental changes to memory system architecture, and assume fully-trusted modules. We propose {\em \papername{}}, a low-cost RAP that targets direct-attached memories, like DDRx. \papername{} avoids memory-side data authentication, and thus, only adds a small amount of logic to memory components and does not change the underlying DDR protocol, making it practical for widespread adoption.
\ignore{We develop \papername{} for both trusted and untrusted memory modules and discuss associated trade-offs.}
In contrast to prior mutual authentication proposals, which require trusting the entire memory module, \papername{} targets untrusted modules by placing its limited security logic on the DRAM die (or package) of the ECC chip.
Our evaluation shows that \papername{} performs within 1\% of an encryption-only memory without RAP and that \papername{} provides 18.8\% and 7.8\% average performance improvements (up to 190.4\% and 24.8\%) relative to a 64-ary integrity tree and an authenticated channel, respectively.

\end{abstract}

\begin{IEEEkeywords}
Memory security, Replay attacks, Memory integrity
\end{IEEEkeywords}

%% file: text/introduction.tex
\section{Introduction}
Trusted data-center infrastructure is crucial for users to move their applications and data to the cloud. One risk is attacks on main memory that have been demonstrated for accessing private data~\cite{halderman2009lest, kwong2020rambleed,trikalinou2017taking,lee2020off} and even for taking over entire servers~\cite{kim2014flipping,seaborn2015exploiting}. To mitigate against main memory vulnerabilities and physical attacks, \textit{trusted execution environments} (TEE), such as Intel \revision{Software Guard Extensions (SGX)}~\cite{gueron2016memory}, provide secure off-chip memory that ensures data confidentiality and integrity. 

Securing memory incurs application slowdown because each memory access requires additional security metadata accesses. 
%For confidentiality, data is encrypted, possibly using \emph{counter-mode encryption} in which each cache-line is associated with an encryption counter to make each encrypted data block temporally unique. 
Of particular interest to this paper is that for integrity protection, each data block is guarded by a cryptographic \emph{message authentication code} (MAC), which is stored with the data in the memory. The processor has to fetch the stored MAC to verify data integrity. In this paper, we focus solely on reducing the memory integrity overheads and rely on unmodified prevalent confidentiality  schemes~\cite{gueron2016memory,inteltme,amdsev}.%\revision{and discuss how our proposal works with each.} %Issues that are the concern of the confidentiality scheme (e.g., side-channels).}

%Although MACs provide strong detection of arbitrary data tampering, they do not offer \textit{complete} integrity protection as they are vulnerable to \textit{replay attacks}. 
The MAC itself must also be protected to provide \emph{complete} integrity guarantees and prevent \emph{replay attacks}. In a replay attack, the attacker bypasses the integrity verification by replaying an \textit{older} pair of data and its MAC (e.g., a 72-byte tuple for 64-byte data and an 8-byte MAC). This pair appears to be correct on the processor, however, it is stale and may corrupt execution. For \emph{replay attack protection} (RAP), secure processors may create an integrity tree over the MACs or over the encryption counters~\cite{rogers2007using, gueron2016memory}. The processor traverses the tree from the leaf to the root to verify the integrity and freshness of the data or counters. The root of the tree is always stored on chip and cannot be tampered with. This integrity tree increases memory bandwidth pressure and access latency as it requires several additional accesses to traverse.

%\ignore{
\placetextbox{.7}{0.085}{\fbox{
\begin{minipage}{.47\textwidth}
\large{This work has been submitted to the IEEE for possible publication. Copyright may be transferred without notice, after which this version may no longer be accessible.}
\end{minipage}
}
}
%}

While prior work has proposed techniques to lower the cost of MACs~\cite{saileshwar2018synergy, fakhrzadehgan2014safeguard} and other security metadata~\cite{yan2006improving}, integrity trees continue to limit scalability and performance of secure memories. This is because the tree traversal overhead is proportional to its size and height, which depends on the protected memory size. Applications with large memory footprints experience a significant slowdown due to either expensive tree walks or the extra data movement caused by the numerous page faults required to manage the small efficiently-protected memory space afforded by small integrity trees~\cite{taassori2018vault}. 

Prior work, such as compact high-arity trees~\cite{taassori2018vault, saileshwar2018morphable}, has had only limited success in addressing this harsh tradeoff between per-access integrity tree overhead and a small protected memory.
This crucial limitation of integrity trees continues to be a major obstacle to widespread commercial adoption of complete memory protection. For example, while new products have extended memory encryption to the entire memory space (e.g., Total Memory Encryption (TME)~\cite{inteltme} and Secure Encrypted Virtualization (SEV)~\cite{amdsev}), replay attack protection is either missing, or is restricted to only a small portion of memory (e.g., 96MB for Intel SGX and a small portion of memory in Apple's Secure Enclave Processor~\cite{apple_enclave}).

We propose {\em \papername{}} to protect the DDR interface against practical replay attacks at much lower cost than current industrial and academic approaches. \papername{} uses a narrow secure channel to \emph{encrypt} the MAC (E-MAC)
and protect it while data is transferred between the processor and memory. This prevents an attacker from replaying a stale \emph{(Data, MAC)} pair as the plain-text MAC is not observable. The channel counters are not stored and are incremented at each memory transaction, making E-MACs temporally unique such that an E-MAC is never repeated with its data. MACs are stored un-encrypted in memory, protecting the integrity of the data at rest. \papername{} performs MAC verification \textit{only} on the processor.

The E-MACs fully guarantee data integrity, but are vulnerable to a stale-data attack where the attacker manipulates the command and address signals to force a memory write to not reach its destination address. The old \emph{(Data, MAC)} are returned when that address is read again, providing a stale pair. We protect against such attacks by introducing \revision{\emph{encrypted write cyclic redundancy code (CRC)}} that extends the \revision{\emph{extended write CRC (eWCRC)}} approach of \revision{All-Inclusive ECC (AI-ECC)}~\cite{AIECC} to allow the memory device to identify mismatched addresses and data before performing the write, thus detecting any tampering. Like E-MACs, we encrypt the eWCRC to both prevent an attacker from choosing values that can still pass the non-cryptographic CRC check and to prevent new information leakage.

\ignore{
To cover physical attacks on the memory module, we develop \papername{} for both trusted and untrusted DIMMs, and discuss the trade-offs between these two threat models. In the \emph{trusted} memory-modules threat model, attackers may not modify or access components within a DIMM. Common inspection techniques help avoid the deployment of malicious modules. In the \emph{untrusted} module threat model, a stronger attacker that can tap or tamper with the on-DIMM components is assumed. To mitigate on-DIMM attacks, we place \papername{}'s security logic in some of the DRAM devices (the ECC chips). While this approach enhances security, implementing the security logic on the DRAM die makes it more costly. However, such logic is feasible considering the advancements in processing-in-memory technologies demonstrated by recent industrial prototypes~\cite{upmem_case_study, fim-dimm, skhynix_pim, pim_samsung_isca21}.
}

{\em Mutually authenticated} channels between the processor and memory have also been proposed to defeat replay attacks without an integrity tree. InvisiMem~\cite{aga2017invisimem} applies this to the packetized protocol of the Hybrid Memory Cube (HMC)~\cite{hmc_gen2}. However, direct adaptation of InvisiMem to DDRx \revision{dual in-line memory modules (DIMM)} is impractical. First, the security guarantees of InvisiMem (and any mutual authenticated channel) require that the entire \revision{DIMM} be trusted. This is acceptable for the logic and memory layers in an HMC, but does not hold true for commodity DIMM-style modules that comprise multiple discrete components. One could extend InvisiMem's trusted computing base (TCB) to include the entire module, however, this leaves the system vulnerable to physical attacks on the DIMM. Moreover, mutual authentication for DDRx requires fundamental changes as DDRx is not packetized, has strict standardized timing parameters, and commodity DIMMs do not have a centralized data buffer in which mutual authentication can be computed (Section~\ref{sec:compare_active_memories}). %for two reasons. First, memory-side integrity verification adds extra latency on the memory-access critical path. Second, memory-side MAC generation requires gathering the data in a \textit{centralized} data buffer on the memory module. Centralized data buffers have been abandoned by modern DIMM designs because they limit the data transfer rate (Section~\ref{sec:compare_active_memories}). By eliminating the memory-side MAC verification in \papername{}, we neither require a centralized data buffer nor change DDRx timings parameters. 
%\textit{ Due to high overheads of integrity trees and limitations of mutual authentication, for architecting secure large capacity memories, the future system design will need to either (a) accept significant slowdown or (b) give up on RAP altogether.}

We develop \papername{} to overcome the challenges of providing a low-cost and scalable RAP for commodity DDRx modules. In contrast with prior work~\cite{aga2017invisimem}, which its successful adoption \emph{requires} trusting the entire memory module, \papername{} can be easily tailored for untrusted DIMMs (as well as trusted) with negligible performance overhead, eliminating the vulnerability to on-DIMM physical attacks and malicious units. %In the \emph{trusted} memory modules threat model, attackers may not modify or access components within a DIMM. In the \emph{untrusted} module threat model, a stronger attacker that can tap or tamper with the on-DIMM components is assumed. 
To this end, we place \papername{}'s limited security logic in some of the DRAM \revision{chips} (the ECC chips). While implementing this logic on the DRAM die is costly, it is practical considering advancements in logic-in-memory technologies demonstrated by DDR5 on-die ECC and recent industrial processing-in-memory prototypes~\cite{upmem_case_study, fim-dimm, skhynix_pim, pim_samsung_isca21}. We anticipate this to be a boon to memory vendors as the market for server memory is large, as well as to processor vendors who can offer highly-secure memory with less overhead.

Overall, this paper makes the following contributions:

\begin{itemize}[itemsep=0.0em]
    \setlength{\parskip}{0pt}
    \item We analyze different replay attack scenarios and observe that replay attacks can be mitigated by protecting only the MACs as they traverse the memory channel.
    
    \item We propose \papername{}, a low-cost replay attack protection mechanism for the DDRx standard. \papername{} uses dedicated encryption units to encrypt MACs, protecting them on the bus, and synchronized channel encryption counters to protect against data-at-rest attacks.
    %\mattan{I think it's OK not to mention eWCRC, but just pointing out that it isn't mentioned in the summary.}
    
    \item To protect against on-DIMM vulnerabilities, we develop \papername{} for untrusted \revision{DIMMs}, including address manipulation and man-in-the-middle attacks. We further discuss how \papername{} is compatible with trusted DIMMs as well.
    
    \item \papername{} enables low-overhead integrity protection while supporting both counter-mode encryption and recent commercial approaches that forego counters for the AES-XTS encryption scheme~\cite{inteltme,amdsev}. We show that AES-XTS provides a substantial performance boost over counter-mode encryption and is compatible with \papername{} but not with state-of-the-art integrity-tree designs.

    \item We evaluate \papername{} and show that it provides 18.8\% and 7.8\% average performance improvements (up to 190.4\% and 24.8\%) relative to a 64-ary integrity tree and an authenticated channel based on InvisiMem, respectively. \revision{\papername{} performs within 1\% and 3\% of encrypt-only memories with AES-XTS and AES-CNT, respectively.}

    %(\makeblue{which its adaptation to DDRx DIMMs does not address on-DIMM vulnerabilities and is not practical for DDRx modules})
    %\mattan{Is it fair to call this InvisiMem or should we say something like "our careful adaptation of InvisiMem to DDRx DIMMs"? -- FIXED}
\end{itemize}

%% file: text/background.tex
\ignore{
2.1 basic secure memory 
- security: integrity + confidentiality
- conf. is fixed 
- intgirty is fixed, if not care replay attack

2.2. threat model with replay attacks
- threat model (for replay attacks)
- define replay attacks
- how replay attack can happen
- talk about address corruption (no real data replay, but stale data in the location)
- same ordering for all attack-defense across the paper

2.3. current defenses
- integrity tree
- mutual
}
\section{Background \& Motivation}\label{sec:background}
\subsection{Threat Model}\label{sec:threat-model}
We consider a threat model similar to SGX~\cite{gueron2016memory,costan2016intel}. The software that runs in the secure environment (e.g., the enclave) is the only software part of the Trusted Computing Base (TCB). Other processes (including the OS and the Hypervisor) are untrusted and are restricted with a hardware-based isolation. An adversary can perform passive (eavesdropping on application information) or active (tampering with the data) physical attacks. The processor chip is part of the TCB and cannot be tampered with. The attacker can target any off-chip component, including the memory bus and DIMMs.

We consider a modern DDR4/5 module architecture (DIMM)\ignore{, as shown in Figure~\ref{fig:modern_dimm},} to cover the attack surface of a memory module. A memory module is composed of several DRAM chips. Each chip has a narrow interface (e.g., 4, 8, or 16 bits). To create a wider data bus, multiple chips are organized in groups called \emph{ranks}, all operating in lockstep within a rank. A module can have multiple ranks for higher capacity.

\ignore{
\begin{figure}[ht]
    \centering
    %\vspace*{-2mm}
    \includegraphics[width=\columnwidth]{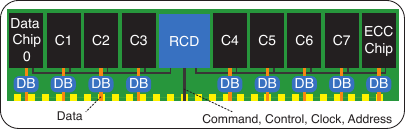}
    \caption{A modern LRDIMM architecture~\cite{idt_lrdimm, rambus_ddr5_db}.}
    \label{fig:modern_dimm}
    %\vspace*{-3mm}
\end{figure}
}

The large number of memory \revision{chips} on a high-capacity DIMM increases the capacitive load on the memory bus and the module's interconnects, which adversely impacts signal stability and integrity. To mitigate this problem, industry has adopted \textit{registered DIMMs (RDIMM)} and \textit{load-reduced DIMMs (LRDIMM)}. In these designs, the I/O signals to each of the DRAM \revision{chips} are decoupled by adding extra buffer chips to the module~\cite{idt_lrdimm, rambus_ddr5_db}. Buffer chips include a single centralized registered clock driver (RCD) chip for the command, control, clock, and address (CCCA) signals, and several distributed data buffers (DB) for buffering the data pins. Whereas RDIMMs only have the RCD to buffer the CCCA, LRDIMMs have both RCD and distributed DBs to buffer both CCCA and data.

\ignore{
While the attacker can target the interconnects on the DIMM, we consider physical attacks on internal circuits of an Integrated Circuit (IC) such as the processor chip, DBs, RCD, and the DRAM devices out-of-scope because these circuits are at micron/nanometer scale, making the attack way too expensive~\cite{suh2005design, aga2017invisimem}. We do not consider Address/Command traffic, bus utilization, power, and electromagnetic side-channels.
}

In line with prior work~\cite{suh2005design, aga2017invisimem}, we consider attacks that target the interconnects on the DIMM, but keep  physical attacks on internal circuits within a package, such as the processor chip, DBs, RCD, and the DRAM \revision{chips} out of scope. In-package attacks are significantly harder to perform as they require successfully desoldering packages, removing different transistor layers to reach the target cells or connections, and tapping circuits that are at micron/nanometer scale while maintaining high-performance operation within a running system.\ignore{Such attacks can be destructive to the device, and require special equipment and expertise that makes them way too expensive~\cite{suh2005design, aga2017invisimem}.} We do not consider address/command traffic, bus utilization, power, and electromagnetic side-channels because these are confidentiality issues. This is out of the scope of this paper as our focus is integrity protection and our approach does not affect the confidentiality mechanisms and \papername{} does not open additional side channels\revision{ (see Section~\ref{sec:side-channels})}.

%\vspace*{-3mm}
\subsection{Secure Memory Basics}\label{sec:secure_memory_basics}

Ensuring the off-chip data security has two aspects: confidentiality and integrity. Confidentiality is needed to protect data \emph{privacy}. Integrity is needed to protect the \emph{correctness} of the off-chip data, i.e., ensuring that data has been indeed written by the trusted software running on the trusted processor and has not been modified by an adversary in the interim.

\vspace{0.05 in}
\noindent{\bf Data Confidentiality.}
\vspace{0.01 in}
Secure processors use \emph{encryption} to ensure data confidentiality. Intel SGX~\cite{gueron2016memory} uses counter-mode encryption, in which each cache-line is associated with an encryption counter that is stored off-chip. Recent products (e.g., Intel TME~\cite{inteltme} and AMD SEV~\cite{amdsev}) have managed to extend memory encryption to the entire memory space by adopting low-cost XOR-Encrypt-XOR (XEX) encryption mode (e.g., AES-XTS)~\cite{wilke2020sevurity}, omitting the encryption-counter storage and memory bandwidth overheads.

\vspace{0.05 in}
\noindent{\bf Data Integrity.}
\vspace{0.01 in}
To protect data integrity, each cache-line is guarded by a\ignore{ cryptographic hash, such as a} \emph{message authentication code} (MAC), which can detect arbitrary data modifications. The MACs are stored with the data in memory and need to be fetched to verify data integrity, which incurs storage and memory bandwidth overheads. To provide low-cost integrity protection, recent products (e.g., Intel TDX~\cite{inteltdx, inteltdx_whitepaper}) and academic proposals (e.g., SafeGuard~\cite{fakhrzadehgan2014safeguard}) place both MAC and \revision{error correction code (ECC)} in the ECC chips and transfer them using the ECC portion of the bus. This eliminates the storage and bandwidth overheads of the MACs while maintaining ECC protection~\cite{fakhrzadehgan2014safeguard}.
%\mattan{Should we just make this generic and include both SafeGuard and TDX in one sentence? Why separate them?}

In this paper, we consider a baseline system equipped with similar low-cost confidentiality and integrity mechanisms. Unfortunately, MACs alone cannot provide \emph{complete} integrity protection as they must also be protected to prevent \emph{replay attacks}. %In a replay attack, the attacker bypasses the integrity verification by replaying an \textit{older} pair of data and its MAC. 
In the next section, we discuss how a replay attack is performed and explain current mitigation techniques.

\subsection{Replay Attacks \& Defenses}\label{sec:replay-attack}

In this section, we formally define replay attacks and describe how they can bypass integrity checks. We explain which data is vulnerable to replay attacks and discuss the existing mitigations.

\ignore{\subsubsection{How to Perform a Replay Attack?}}
\vspace{0.05 in}
\noindent{\it 1) How to Perform a Replay Attack?}
\vspace{0.01 in}

\begin{figure}[ht]
    \centering
    \vspace*{-3mm} 
    \hspace*{-3mm}
    \includegraphics[width=0.75\columnwidth]{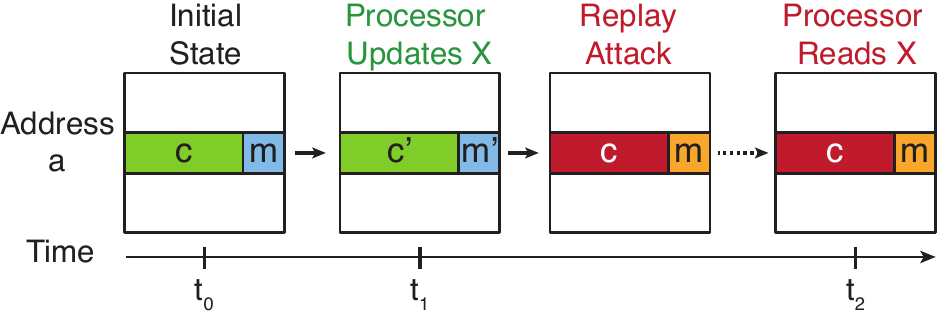}
    %\vspace*{-3mm}
    \caption{Logical view of replay attack on address $a$.}
    \label{fig:replay-attack}
    %\vspace*{-2mm}
\end{figure}

A replay attack can bypass integrity verification if it does not result in a MAC mismatch. Any corruption in the data or its MAC causes an integrity verification failure with sufficiently high probability, except when both data and its MAC are replayed at once. In other words, if $(c, m)^{a}_{t_{0}}$ is the state of the cache-line $c$ at address $a$ with its MAC $m$ at time $t_{0}$, and $(c, m)^{a}_{t_{1}}$ is the state at time $t_{1} > t_{0}$, overwriting the tuple $(c, m)^{a}_{t_{1}}$ with $(c, m)^{a}_{t_{0}}$ would pass the MAC verification of the line $c$. Note that it is important to replay the tuple to the \emph{same address} since physical addresses are included in the MAC~\cite{gueron2016memory,yan2006improving}. Thus, the attacker has to precisely track memory addresses, memoize changes to a specific location over time, and precisely replay a $(\textrm{\textit{Data}}, \textrm{\textit{MAC}})$ tuple to avoid signaling an integrity violation. Figure~\ref{fig:replay-attack} depicts logical view of a replay attack on address $a$. Integrity verification passes at time $t_{2}$. 
%\mattan{Are MACs really 8B? Is that a requirement? I mean, clearly not :-). Perhaps change this a bit? -- Ali: Not really, even in SGX it's 56 bits. But, I think getting to such details might be distracting. We can maybe make it a range 4-8B MAC.}

\ignore{\subsubsection{Which Data is Vulnerable to a Replay Attack?}}
\vspace{0.05 in}
\noindent{\it 2) Which Data is Vulnerable to a Replay Attack?}
\vspace{0.01 in}

We categorize different types of replay attacks based on whether they are done on data at rest or data in motion.

\vspace{0.05 in}
\noindent{\bf Data at Rest.}
\vspace{0.01 in}
Off-chip data {\em at rest} is the data that is stored in memory that the application is not currently operating on. TEEs protect secure environments (enclaves) using hardware-based isolation mechanisms~\cite{costan2016intel} so that different processes (including the OS) cannot access each other's data. Thus, a software-based replay attack could not succeed. %\mattan{Unlikely or impossible?}

The attacker could attempt to replay the data \textit{indirectly} by inducing bit-flips (e.g., via Row-Hammer~\cite{kim2014flipping} or causing soft errors), however, we consider this type of attack impractical. Theoretically, it is not impossible to perform a replay attack by bit-flips, however, the likelihood of success would be extremely low as the attack needs to flip enough bits such that $(\textrm{\textit{Data}}, \textrm{\textit{MAC}})$ match precisely. All demonstrated Row-Hammer attacks only induce a few bit-flips per-line (fewer than 10)~\cite{kim2020revisiting, half-double}. We do not know of any real-world replay attacks that have occurred using these means.

The attacker can target data at rest using {\em DIMM substitution}. The attacker keeps a version of the data by removing the DIMM and replaying the application state by plugging in that DIMM later. This attack relies on the {\em data remanence} effect~\cite{halderman2009lest}.\footnote{Theoretically, the attacker can detach the memory chips from the module PCB and use a different PCB to replay the data. While this attack is possible on a module assumed to be trusted with no on-DIMM protection, \papername{} defeats this attack, as we describe in Section~\ref{sec:tcb}.}

\vspace{0.05 in}
\noindent{\bf Data in Motion.}
\vspace{0.01 in}
Off-chip data {\em in motion} is data that is being transferred between the memory and the processor, such as an LLC fill or write-back. %The attacker can perform a replay attack on data in motion on either the read or write path of the data. In this case, 
The replay attack is a {\em Man-In-The-Middle} attack, where the attacker either interposes traffic on the bus between the processor and the memory module or uses a malicious DIMM that is capable of analyzing and intercepting on-DIMM interconnect (e.g., via a trojan).

\ignore{
\begin{tcolorbox}
\vspace{-1mm} {\bf Insight-1:}
 snapshots, 
Practical replay attacks \emph{require} bringing data into motion.
\end{tcolorbox}
}

\ignore{\subsubsection{Current Defenses Against Replay Attacks}\label{sec:current-defense}}
\vspace{0.05 in}
\noindent{\it 3) Current Defenses Against Replay Attacks}\label{sec:current-defense}
\vspace{0.01 in}

\vspace{0.05 in}
\noindent{\bf Integrity Trees.}
\vspace{0.01 in}
To defeat replay attacks, secure processors create an integrity tree over the MACs~\cite{gassend2003caches} or over the encryption counters~\cite{rogers2007using, gueron2016memory}. The processor traverses the tree from the leaf to the root to verify the integrity and freshness of the data or counters. The root is always stored on-chip and cannot be tampered with. For more details on tree designs and traversal refer to prior work~\cite{gueron2016memory,costan2016intel,rogers2007using,yan2006improving,taassori2018vault,saileshwar2018morphable}. 

\ignore{
\begin{figure}[ht]
    \centering
    %\vspace*{-2mm}
    \includegraphics[width=0.9\columnwidth]{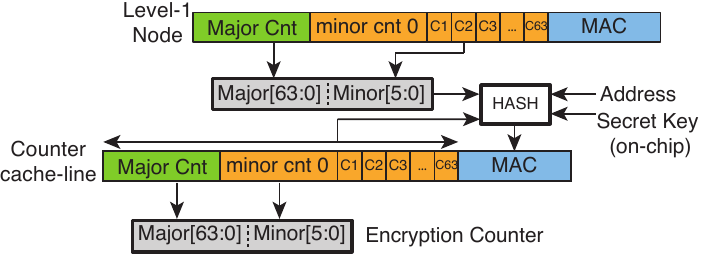}
    \caption{64-ary split-counter tree organization~\cite{yan2006improving,taassori2018vault, saileshwar2018morphable}.}
    %\vspace*{-2mm}
    \label{fig:integrity-tree-node-split-counter}
\end{figure}
}

\vspace{0.05 in}
\noindent{\bf Mutual Authentication.}
\vspace{0.01 in}
Mutually authenticated channels between the processor and memory defeat replay attacks without an integrity tree. Each memory transaction on the bus is protected using a unique and dedicated per-transaction MAC ($\textrm{\textit{MAC}}_{t}$). A $\textrm{\textit{MAC}}_{t}$ is generated with a transaction secret key ($K_{t}$), the data, and a non-repeating nonce, such as a counter ($C_{t}$), which both ends (the processor and the DIMM) are equipped with. $C_{t}$ is incremented at each transaction and always has the same value on both ends. At each transaction, the sender (processor on writes and DIMM on reads), uses $C_{t}$ to compute $\textrm{\textit{MAC}}_{t}$ for the transmitted data, i.e., $\textrm{\textit{MAC}}_{t} = H_{K_{t}}(Data, C_{t})$.  On the receiving end, the MAC is recomputed and compared against the received $\textrm{\textit{MAC}}_{t}$ to verify data integrity and freshness before being stored in memory or used in the processor. Because $C_{t}$ is unique, replay attacks are detected.

\subsection{Goal: Practical Replay Attack Protection}
Ideally, replay attack protection should be scalable, low-cost, and provide complete protection. Integrity trees do not scale to large capacity memories. Prior attempts that create a mutually authenticated channel on the memory bus require trusting the entire module and use packetized protocols, which are not applicable to modern DDRx DIMM design constraints. Our goal is to develop a low-cost solution that meets these requirements. Our aim is to make this solution practical for widespread adoption, and applicable to contemporary DIMMs without modifying the underlying memory protocol.

%% file: text/sentry.tex
\section{\papername{}: Low-Cost Replay Attack Protection}\label{sec:proposed_method}

\papername{} is based on the insight that integrity can be provided by blocking replay attacks on the bus. In brief, \papername{} creates a replay-protected channel on the memory bus by modulating the MACs for the data that is in transfer on the bus, as summarized in Section~\ref{sec:idea_core}.
%and how it protects data-in-motion using encrypted MACs. %Section~\ref{sec:enc-count-protection} describes how \papername{} can be extended to protect the encryption counters integrity.
Section~\ref{sec:cmd_addr_integrity} describes how \papername{} protects against attacks on CCCA signals that feed stale data to the processor. We discuss how \papername{} protects from DIMM-substitution attacks in Section~\ref{sec:dimm_subs}. We describe \papername{}'s TCB in Section~\ref{sec:tcb}. Section~\ref{sec:remote-attestation} describes system initialization and the attestation process.

%The non-repeating nature of these encryption counters makes E-MACs temporally unique. Thus, if an attacker tries to tamper or replay an older \textit{(Data, MAC)} on the bus, the E-MAC will be decrypted to an incorrect value. To protect the data at rest against tampering (e.g., Row-Hammer) and provide complete memory integrity, decrypted MAC is stored with the data in the memory.

\subsection{Replay-Protected Bus Using E-MACs}\label{sec:idea_core}
\ignore{ 
To protect data at rest against tampering, we cannot discard the MACs, and instead, we need to store the MACs alongside the data in the storage device and verify it on the subsequent accesses, similar to SGX. However, $\textrm{\textit{MAC}}_{t}$ is generated using $C_{t}$, which is incremented at each data transfer and is specific to each transaction. Thus, only storing $\textrm{\textit{MAC}}_{t}$ is not sufficient since without having the right counter ($C_{t}$), MAC re-computation and verification would not be feasible. Storing $C_{t}$ with $\textrm{\textit{MAC}}_{t}$ solves this problem. However, considering 64-bit counters, this incurs 25\% storage overhead (in addition to the existing MAC and the encryption counters), as we also need to protect the $C_{t}$.
}

Although a mutually authenticated bus (Section~\ref{sec:current-defense}) protects the integrity of data in motion, it is not sufficient to protect integrity of data at rest. We need to store a MAC with the data in memory to verify its correctness on subsequent accesses (as in SGX/TDX). However, both ($\textrm{\textit{MAC}}_{t}$) and ($C_{t}$) must be stored together for later verification, as ($C_{t}$) is incremented dynamically with every direction. This has a very high 25\% total storage overhead for 64-bit counters and MACs.
%, as we also need to protect the $C_{t}$ integrity.

One alternative is to discard $\textrm{\textit{MAC}}_{t}$ and delegate integrity protection of the data at rest to the memory module. On each data write, after verifying $\textrm{\textit{MAC}}_{t}$, the memory module generates a new MAC and stores it with the data. On reads, the memory module first performs a MAC verification, and if this verification passes, it can then generate $\textrm{\textit{MAC}}_{t}$ and transmit the data on the bus. InvisiMem~\cite{aga2017invisimem} uses this technique, however, adapting this approach requires trusting the entire \revision{DIMM} and fundamental changes in the \revision{DDRx} module architecture (see Section~\ref{sec:compare_active_memories}), making it impractical.

\ignore{
\begin{tcolorbox}
\vspace{-1mm} {\bf Insight-2:} {\em Memory-side} integrity verification is the main reason prior mutual authentication proposals are unfit for DDRx modules (see Section~\ref{sec:bad_invisimem}). \vspace{-2mm}
\end{tcolorbox}
}

\begin{figure}[ht]
    \centering
    %\vspace*{-3mm}
    \includegraphics[width=0.6\linewidth]{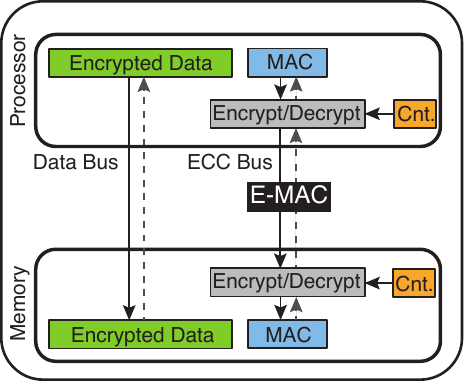}
    %\caption{\papername{} encrypts MACs to make them temporally unique and capable of detecting replay attacks.}
    \caption{\papername{} overview.}
    \label{fig:idea-emac}
    %\vspace*{-3mm}
\end{figure}

Given that replay attacks require bringing data into motion %(Insight-1)
and that replay attack protection in mutual authentication is provided by making $\textrm{\textit{MAC}}_{t}$ {\em temporally unique}, we propose to eliminate memory-side integrity check via MAC encryption.

\ignore{
\begin{tcolorbox}
{\bf Insight-3:} To protect data in motion, we can encrypt the MAC that protects the data at rest using a non-repeating nonce to make it temporally unique.
\end{tcolorbox}
}

\papername{} uses \textit{\underline{E}ncrypted \underline{MAC}s (E-MACs)} to protect the bus. On a data write, the processor's memory encryption engine generates a MAC using $\textrm{\textit{MAC}} = H_{k}(data, addr)$. However, before transferring this MAC on the bus, it is first encrypted to generate the E-MAC. Figure~\ref{fig:idea-emac} shows an overview of \papername{} as it uses MACs to protect data at rest and repurposes them for protecting the data in motion. To generate the E-MAC, we XOR the MAC with a \emph{one time pad} ($\textrm{\textit{OTP}}_{t}$) generated using the transaction counter $C_{t}$.\ignore{, as shown in Figure~\ref{fig:per-transaction}.} This effectively makes the MAC temporally unique and capable of detecting memory bus replay attacks (same as $\textrm{\textit{MAC}}_{t}$ in mutual authentication). The per-rank transaction counter is incremented at both the memory controller and memory module.

\ignore{
\begin{figure}[ht]
    \centering
    %\vspace*{-2mm}
    \includegraphics[width=0.8\columnwidth]{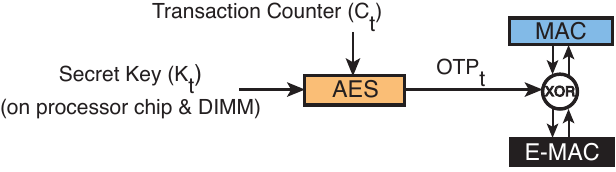}
    \caption{Using encryption to get temporally-unique MAC.}
    \label{fig:per-transaction}
    \vspace*{-4mm}
\end{figure}
}

In \papername{}, \textit{only} the processor performs integrity verification, and as a result, there is an important difference between how the processor and the DIMM use E-MACs. On each receiving end, we first XOR the E-MAC with the $\textrm{\textit{OTP}}_{t}$ to decrypt it and retrieve the original MAC. On the DIMM, this MAC is not used for verification and is simply stored, and the $C_{t}$ discarded. Since this MAC is not verified on the DIMM, any attack at the time of write will remain undetected until the next read, just as with integrity trees. On the processor, this MAC is used for integrity verification, and a mismatch signals a failure, which could be due to multiple different sources:

\begin{itemize}[leftmargin=7mm,itemsep=0.0em]%parskip=0]
\setlength{\parskip}{0pt}
\item Bit-flips on the data-bus on a data read/write.
\item Bit-flips while data was stored in the memory.
\item Replay attack on the data read.
\item Replay attack on a prior data write.
\end{itemize}

The processor cannot distinguish between different attack types, however, it can detect that {\em an} attack has occurred and that the data has been tampered with, which is what matters for guaranteeing integrity. This is true since tampering with the E-MAC causes a wrong MAC to be computed after it is XOR-ed with the $\textrm{\textit{OTP}}_{t}$. This is also true for write accesses, except the wrong MAC is stored, and its verification is deferred until the next read.

\vspace{0.05 in}
\noindent{\bf Compatibility with On-Die ECC.}
\vspace{0.01 in}
\papername{} provides replay attack protection by encrypting the MACs (E-MACs) to make them temporally unique. Although we have developed \papername{} based on state-of-the-art designs that place MACs in the rank-level ECC to mitigate MAC access overheads~\cite{inteltdx, saileshwar2018synergy, fakhrzadehgan2014safeguard}, MAC encryption is effective regardless of these optimizations.

\subsection{Ensuring Command \& Address Integrity}\label{sec:cmd_addr_integrity}
Attacks that we have considered so far are accomplished by directly targeting the data. However, the attacker can modify the CCCA signals to corrupt data integrity.% This is possible because modifying the sequence of commands and their timings in a way that violates the DRAM timing constraints can put the row buffers and the bit lines in a state where data cannot be reliably read or written. For example, if a DRAM row is already opened and the processor needs to access a different row, before issuing an \textit{Activate} command to open the new row, the memory controller has to first issue a \textit{Precharge} command to close the already-opened row and set the bit-lines voltage to $V_{DD}/2$. Thus, omitting the \textit{Precharge} command results in reading a wrong value, violating data integrity. Fortunately, in an integrity-protected memory, such cases can be easily detected since these violations lead to arbitrary data corruption that MACs can detect. However, we identify the following attack scenario in which the attacker can tamper with the address bus to form a replay attack.

\vspace{0.05 in}
\noindent{\bf Attack Scenario.}
\vspace{0.01 in}
In \papername{}, any data corruption (including replay attacks) that happens at the time of write is not detected immediately and is deferred until the subsequent read. This method is safe only if the corrupted data is written \textit{in place}, overriding the previous version of the $(\textrm{\textit{Data}}, \textrm{\textit{MAC}})$ tuple. However, if the write is redirected to a different memory location, the stale $(Data, \textrm{\textit{MAC}})$ will remain in place, and the processor cannot tell that it is out of date. 

\begin{figure}[ht]
    \centering
    %\vspace*{-2mm}
    \includegraphics[width=.9\linewidth]{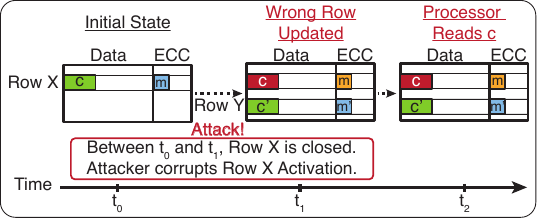}
    %\includegraphics[width=\linewidth]{graphs/address_attack_horiz.pdf}
    %\vspace*{-1mm}
    \caption{Performing replay attack by corrupting the address bus. %(a) 
    The attacker corrupts the row address of the Activate command when the processor is writing to cache-line $c$. 
    %(b) The attacker corrupts the column address of the Write.
    }
    \label{fig:address_attack}
    \vspace*{-3mm}
\end{figure}

Figure~\ref{fig:address_attack} shows an example in which the attacker creates such a scenario by corrupting the write address. Assume the processor reads the cache-line $c$ at $t_0$, updates it to a new value $c'$ at $t_1$, and attempts to read it at a later time $t_2$. In this case, between the time that the processor initially reads $c$ and when it wants to write $c'$, the DRAM row that $c$ belongs to (row $X$) is closed. The memory controller has to first open row $X$, however, the attacker intercepts the \textit{Activate} command and changes the row address to a different row (row $Y$). As a result, when the processor performs the write, $c'$ will be written to the wrong row, leaving the original location with the stale $(\textrm{\textit{Data}}, \textrm{\textit{MAC}})$. When the processor attempts to read the data at $t_2$, it opens row $X$ reading the stale tuple, which passes MAC verification, completing the replay attack cycle. %Figure~\ref{fig:address_attack}b is a similar attack, however,

In a similar attack, instead of corrupting the row address, the attacker can change the column address, writing $c'$ to a different column in the original row. Note that if the processor ever reads the location that the attacker has redirected the writes to (i.e., row $Y$ or the wrong column), \papername{} detects the attack as the line address is included in the MAC~\cite{gueron2016memory}. However, the attacker can orchestrate the attack in a way that remains undetected. Alternatively, the attacker can simply drop the write request instead of redirecting it to a different location. However, \papername{} can detect this case since dropping a request will cause a $C_{t}$ mismatch between the processor and memory.

\begin{figure}[ht]
    \centering
    %\vspace*{-2mm}
    \includegraphics[width=.9\linewidth]{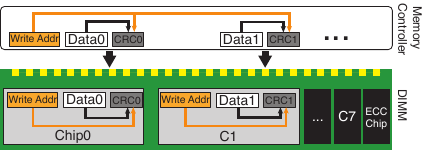}
    \caption{AI-ECC's eWCRC~\cite{AIECC}. The memory controller encodes the write address with the data in the WCRC. Each chip uses the address and data to verify the transaction.}
    \label{fig:aiecc_ewcrc}
    \vspace*{-2mm}
\end{figure}

\vspace{0.05 in}
\noindent{\bf All-Inclusive ECC (AI-ECC)~\cite{AIECC}.}
\vspace{0.01 in}
CCCA corruption can also happen from naturally occurring faults. AI-ECC mitigates these errors by extending the existing reliability measures of contemporary DRAM \revision{chips} that protect data integrity to also protect the CCCA signals. For early-detection of write data transmission errors, DRAM \revision{chips} use \textit{write cyclic redundancy codes (WCRC)}~\cite{micron_ddr4,micron_ddr5_datasheet}, which are generated over the data transmitted to each chip. Enabling WCRC requires increasing the write burst length from 8 to 10 in DDR4 (16 to 18 in DDR5), in which the WCRC is transmitted to each \revision{chip} over the last two beats (i.e., 16-bit WCRC with x8 device). Before storing the data, each DRAM \revision{chip} internally recomputes the WCRC to make sure transmission was error-free. AI-ECC\footnote{AI-ECC makes additional contributions to protect command and clock signals, however, these cases are detectable in an integrity-protected memory, and we do not further discuss them here.} proposes \textit{extended write CRC (eWCRC)}, which enhances the WCRC to also include the \textit{rank, bank, row,} and \textit{column} address of the write to protect address bus integrity, as shown in Figure~\ref{fig:aiecc_ewcrc}.

%\vspace{0.05 in}
%\noindent{\bf \papername{} with AI-ECC's eWCRC.}
%\vspace{0.01 in}

\vspace{0.05 in}
\noindent{\bf \papername{} with Encrypted eWCRC.}
\vspace{0.01 in}
\papername{} defeats stale-data attacks from misdirected writes by enabling eWCRC and encrypting it, similarly to the E-MAC. The eWCRC of the ECC chip is generated before encrypting the MAC, and it is verified in the ECC chip after decryption of the E-MAC. However, because the eWCRC is not a cryptographic hash, the adversary can target specific bits in the message such that the corrupted eWCRC check would incorrectly pass. This is true even if the eWCRC is encrypted with the $\textrm{\textit{OTP}}_t$ used for the E-MAC. \papername{} therefore uses a separate $\textrm{\textit{OTP}}_t^w$ for write commands that uses the same key and transaction counter, but also includes the address used in eWCRC. This ensures that any corruptions to the address would flip numerous bits in the message and the eWCRC would detect the corruption. This approach increases the write latency because generating the $\textrm{\textit{OTP}}_t^w$ only starts after the write command is sent to the \papername{} DRAM \revision{chip} and takes longer than \texttt{tWCL}. Importantly, the read latency is unaffected because the processor performs MAC verification. An attempt to induce stale data by dropping a write transaction fails because $C_t$ would not be incremented on the memory side, leading all following reads to fail verification.\cameraready{ Finally, the attacker can potentially avoid updating a memory location by converting a write command to a read (and intercepting the response so the processor is not notified), which does not affect synchronization of the counters, and thus, remains undetectable. This attack can be defeated by simply using only even counter values for reads and odd counter values for writes so that command corruption results in a counter mismatch.}
\ignore{The eWCRC is also encrypted with the $\textrm{\textit{OTP}}_t$ before transmission to prevent any information from leaking through the non-cryptographic CRC computation.} 
%One problem with this method, however, is that the eWCRC may reveal some information about the (un-encrypted) MAC, allowing the attacker to escape the eWCRC-check over time. To avoid this problem, we further encrypt the eWCRC using the remaining bits of the $\textrm{\textit{OTP}}_t$ before transmission. 
\ignore{Note that the \textit{channel} bits are not included in the eWCRC. Thus, the attacker can potentially change the channel bits and keep the remaining address bits the same to redirect a write to a different channel without causing an eWCRC mismatch. However, we assume each channel has a separate memory controller with a distinct transaction counter, and thus, redirecting a write to a different channel would lead to a counter mismatch, allowing \papername{} to detect the attack.} 

%math here: https://docs.google.com/spreadsheets/d/1WgWg6_USRc687AJy7FbW5KP7BK4oZyFTSJaOB3CKZ_U/edit?usp=sharing

\vspace{0.05 in}
\noindent{\bf Security of Encrypted eWCRC.}
\vspace{0.01 in}
Assuming the worst-case bit error rate (BER) of $10^{-16}$ on the CCCA signals that is allowed by the JEDEC standard~\cite{micron_ddr5_datasheet}, channel transmission rate of $3200MTps$\footnote{We use half the DDR data rate for the CCCA signals~\cite{micron_ddr5_datasheet}.}, and 26 CCCA and data signals for an x8 device~\cite{micron_ddr5_datasheet}, we expect to observe one CCCA error every $11.13$ days per memory channel on average. Because the attacker only observes the eWCRC and MACs in their counter-encrypted form, \emph{birthday attacks} on the eWCRC are not possible. In a brute force attack, each attempt has a success rate of $2^{-16}$ with the 16b eWCRC. Thus, even with a success probability of only $50\%$, the attacker must perform at least $4.5\times10^{4}$ attempts. Given that CCCA errors due to natural faults are rare and that a higher than-expected transmission error rate indicates an active attack, it takes 1,385 years to exhaust all trials on a single memory channel. In practice, the BER is much lower than the DDRx standard specifies, in the range of $10^{-22}$ to $10^{-21}$~\cite{AIECC}, increasing the brute force attack duration to 138 million years. Even if the attacker launches a parallel attack on 1,000 nodes that each has 16 memory channels, the attack would still take more than 86,000 years.

\subsection{DIMM-Substitution Attack Protection}\label{sec:dimm_subs}
\vspace{0.05 in}
\noindent{\bf Attack Scenario.}
\vspace{0.01 in}
An adversary can perform a replay attack via DIMM-substitution by taking advantage of the {\em data remanence} effect (Cold-boot Attacks~\cite{halderman2009lest}) to replay a victim application's state across boot or wake-up episodes. The attacker causes the system to crash or go to an idle state (i.e., DRAM self-refresh mode), takes away the DIMM, and freezes it to preserve the application state (and potentially copies it). After reboot or wake-up, the victim process continues execution as usual. At a later time, the attacker forces the system into crash/idle again. In the last step, instead of rebooting from the most recent state, the attacker uses the preserved \textit{old} state, which forces the victim application to redo the \textit{already-performed} computations, completing the replay attack cycle.\ignore{, shown in Figure~\ref{fig:dimm_subs}.

\begin{figure}[h]
    \centering
    \includegraphics[width=0.9\linewidth]{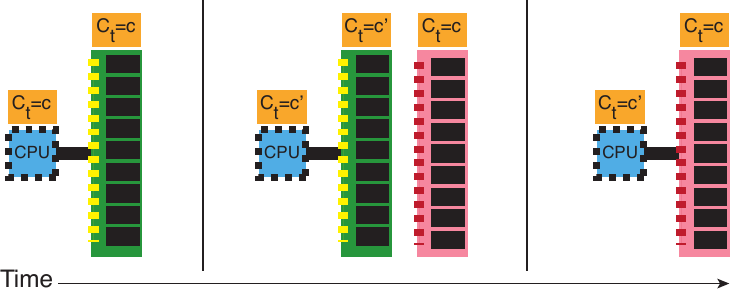}
    \caption{DIMM-Substitution attack overview.}
    \label{fig:dimm_subs}
\end{figure}
}

\vspace{0.05 in}
\noindent{\bf \papername{}'s Efficacy Against the Attack.}
\vspace{0.01 in}
\papername{} defeats this type of attack by using the transaction counters ($C_{t}$). When the attacker tries to wake-up the system using the old state, it is improbable that $C_{t}$ on the DIMM and the processor would match (the likelihood of a match is $\frac{1}{2^{64}}$), causing the $\textrm{\textit{OTP}}_{t}$ on the DIMM and the processor to be different (producing and retrieving different E-MACs), and the attack to fail. With a 64-bit $C_{t}$, we will not observe counter overflow in the system lifetime, as it takes more than 500 years to cause an overflow, even assuming one transaction every nanosecond per rank.
%\revision{Note that, with a 64-bit $C_{t}$, we will not observe counter overflow in the system lifetime. Moreover, because a new key is chosen on every power-up, a $C_t$--$K_t$ pair will not repeat for hundreds of years of operation, even at a pin bit rate of $32$Gb/s.}

\vspace{0.05 in}
\noindent{\bf Non-Adversarial DIMM Replacement.}
\vspace{0.01 in}
It is possible that a DIMM should be replaced for various legitimate reasons (e.g., system upgrade, faulty device) that must be differentiated from an attack. The difference between such cases and a DIMM-substitution attack is that the processor is explicitly notified of the replacement and expects to start from a \textit{clean} state (as opposed to continuing from the previous architectural state in the memory). That is, any prior data in the memory should be discarded, by clearing the memory during boot or DIMM initialization (see Section~\ref{sec:remote-attestation}). %and perform the attestation to coordinate a new a $K_t$ and $C_t$ with the memory module.

\ignore{
\newcommand{\greencheck}{\textcolor{green}{\ding{51}}}
\newcommand{\blackcheck}{\textcolor{black}{\ding{51}}}
\newcommand{\redcross}{\textcolor{red}{\ding{54}}}
\newcommand{\blackcross}{\textcolor{red}{\ding{54}}}
\begin{table*}[ht]
  \centering
  \small
  \caption{Comparison of Integrity Tree and \papername{}.}
  \label{table:tree_vs_idea}
  \begin{tabular}{|c||c||c|c|}
    \hline
    \multirow{2}{*}{Attack Type}                    & \multirow{2}{*}{Integrity Tree}       & \multicolumn{2}{c|}{\textbf{\papername{}}}                   \\\cline{3-4}
                                  &                                       & Trusted DIMM                  & Untrusted DIMM      \\\hline
    Man-in-the-middle (Memory Bus)        & \ding{51}                             & \ding{51}                         & \ding{51}             \\\hline
    Man-in-the-middle (Trojan On-DIMM)           & \ding{51}                             & \redcross                         & \greencheck           \\\hline
    DIMM Substitution & \ding{51}                             & \ding{51}                         & \ding{51}      \\\hline \hline
    Hardware TCB & Processor Chip & Processor Chip + DIMM & Processor Chip + ECC Chip \\\hline \hline
    Performance Overhead (w.r.t encrypt-only) & \% (\%) & \multicolumn{2}{c|}{\% (\%)} \\\hline
  \end{tabular}
\end{table*}
}

\revision{
\subsection{Vulnerability to Side-Channels}\label{sec:side-channels}
\papername{} does not introduce any new side-channels. All of the used cryptographic primitives have constant latency that is independent of the data; the latencies are either hidden from the access critical path or are equally imposed on all accesses. The encryption/decryption of E-MACs does not add any timing variation to reads, as the $\textrm{\textit{OTP}}_{t}$ is always pre-computed independently of transaction timing. The CRC logic also has constant latency. Thus, its extra latency and that of the longer write burst remains indistinguishable among different writes. With \papername{}, writes are slower than reads, however, this does not open a new physical side-channel as the command type and flow of traffic are already observable on the memory bus. From the software perspective, this is not a new side-channel as writes are already slower than reads because the memory controller prioritizes reads.
}

\subsection{Trusted Computing Base for \papername{}}\label{sec:tcb}
The processor chip is the only hardware component in SGX's TCB. We \textbf{must} extend the TCB to include \papername{}'s security logic. This logic includes the secret key register, the encryption units for generating E-MACs, and the attestation logic.

\begin{figure}[h]
    \centering
    %\vspace*{-1mm}
    %\includegraphics[width=\linewidth]{graphs/tcb.pdf}
    \includegraphics[width=.9\linewidth]{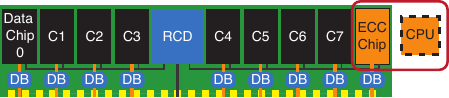}
    \caption{\papername{}'s hardware TCB. 
    %(a) CPU and the DIMM are in the TCB. ECC DB contains the security logic and acts as the DIMM's root-of-trust. (b) 
    CPU and the ECC chip(s) are in the TCB. The ECC chip(s) contains the security logic.}
    \label{fig:tcb}
    %\vspace*{-2mm}
\end{figure}

\ignore{
%\makeblue{To have a baseline that is iso-secure to the prior mutual authentication work~\cite{aga2017invisimem}, we develop \papername{} with two threat models, and discuss the placement of the on-DIMM security logic and the TCB components accordingly. The security logic includes the secret key register, the encryption units for generating E-MACs, and the attestation logic. In both cases, the security logic \textbf{must} be part of the TCB.}

\vspace{0.05 in}
\noindent{\bf 1) TCB with Trusted DIMM:}
\vspace{0.01 in}
We consider this model to mimic the case that InvisiMem's~\cite{aga2017invisimem} security logic is placed in a discrete component on the DIMM (Section~\ref{sec:compare_active_memories}). Similarly, as shown in Figure~\ref{fig:tcb}a, we place \papername{}'s security logic inside the data buffer (DB) of the ECC chip(s). The ECC DB acts as the DIMM's root-of-trust, and after passing attestation and establishing the secure E-MAC channel to the processor. We assume the entire DIMM is in the TCB. Note that an attacker can perform a man-in-the-middle replay attack by tampering with the DIMM interconnects or using a malicious DIMM that contains a hardware \emph{trojan}. These attacks are impractical on 3D-stacked memories as the TSVs between the security logic and the DRAM layers are orders of magnitude smaller than the DIMM interconnects~\cite{aga2017invisimem}. To address these vulnerabilities, we next discuss an untrusted DIMM model.} %We can take additional steps, such as inspecting the module~\cite{mehta2020big}, to assure the DIMM is trojan-free, or

Figure~\ref{fig:tcb} shows \papername{}'s TCB.\footnote{To fairly compare with prior work~\cite{aga2017invisimem}, we discuss \papername{}'s compatibility with trusted memories in Section~\ref{sec:secddr_with_trusted}.} A powerful adversary can initiate the attack from within the DIMM by tapping or tampering with the on-DIMM components or using a malicious DIMM. To defeat these scenarios, we place \papername{}'s security logic within the the ECC chip(s), making the ECC chip(s) part of the TCB. Only trusting the ECC chip(s) is sufficient to detect all active attacks, as tampering with the data chips will cause a MAC mismatch. If the memory module has multiple ranks, the ECC chip(s) in each rank are independent. The processor must establish a separate secure E-MAC channel and use a different transaction counter for each rank.

\ignore{
\vspace{0.05 in}
\noindent{\bf 2) TCB with Untrusted DIMM:}
\vspace{0.01 in}
}

\subsection{Initialization \& Attestation in \papername{}}\label{sec:remote-attestation}
\vspace{0.05 in}
\noindent{\bf Memory Attestation.}
\vspace{0.01 in}
We adopt an attestation protocol similar to prior proposals~\cite{aga2017invisimem, awad2017obfusmem, shafiee2018secure}. The memory manufacturer embeds \emph{endorsement} public-private keys ($\textrm{\textit{EK}}_p$ and $\textrm{\textit{EK}}_s$) in %the ECC DB (for the trusted DIMM) or 
each rank's ECC chip(s).
%(for the untrusted DIMM).
While $EK_p$ is accessible for attestation, $\textrm{\textit{EK}}_s$ never leaves the chip. 

At each power up or DIMM replacement, the processor and each rank use public-key encryption to agree on a new shared key, i.e., the transaction key ($K_{t}$). We use an authenticated key-exchange protocol that is protected against impersonation and man-in-the-middle attacks~\cite{costan2016intel,key-agreement}. For authentication, we use $\textrm{\textit{EK}}_s$ to sign the memory module's key-exchange messages. The \revision{DIMM}'s certificate and $\textrm{\textit{EK}}_p$ should also be shared with the processor to authenticate the key-agreement and to be verified against the DIMM's certificate via a trusted \emph{certificate authority (CA)}. This information can be shared either during the key exchange~\cite{key-agreement}, or it can be manually entered by the \emph{trusted} system-integrator~\cite{awad2017obfusmem}. The CA can be the memory vendor or a third party. \revision{Certificates may be cached in system-encrypted memory and periodically checked against revocation lists}.%A similar approach for CA designation has been proposed in Compute Express Link (CXL)~\cite{cxl_ca}.

After confirming the memory module's identity and sharing $K_{t}$, the processor and memory agree on a common $C_{t}$. The processor chooses an initial counter value for each rank and transfers it to the DIMMs. $C_{t}$ can be shared in plain-text and does not require integrity protection; tampering with the counter results in counter mismatch between the processor and memory, which will be detected through MAC verification failures. We can initialize the counter with a random number, or we can use a non-volatile register for the processor's $C_{t}$ to use monotonically increasing values for the processor lifetime. 

Finally, the processor actively clears memory (writing zeros) to protect from DIMM substitution or replaying stale pre-boot state. Note that attestation is infrequent (at each boot, DIMM power-up, or after non-adversarial DIMM replacement) and only incurs a slight slowdown at that time (seconds). DRAM-timing re-calibration, wake-up from sleep, and other cases where DRAM is already initialized do not require attestation.
%\revision{With a 64-bit $C_t$ and because a new key is chosen on every power-up, a $C_t$--$K_t$ pair will not repeat for hundreds of years of operation even at a pin bit rate of $32$Gb/s.}

\ignore{
%\vspace{-1mm}
\begin{tcolorbox}
{\bf Insight-5:} With a 64-bit $C_t$ and because a new key is chosen on every power-up, a $C_t$--$K_t$ pair will not repeat for hundreds of years of operation (even at a pin-bit rate of $32$Gb/s), defeating DIMM-substitution attacks.
\end{tcolorbox}
%\vspace{-1mm}
}

\vspace{0.05 in}
\noindent{\bf Remote Attestation.}
\vspace{0.01 in}
\papername{}'s DIMM attestation and establishing the secure E-MAC channel on the bus are transparent to the remote attestation of the processor and the software running in an enclave. We can use the same protocol as in SGX to attest the processor, retrieve an enclave's measurement, and create a secure channel to a remote client~\cite{costan2016intel}.

\newcommand{\greencheck} {\large{\textcolor{ForestGreen}{\ding{51}}}}
\newcommand{\blackcheck} {\large{\textcolor{black}      {\ding{51}}}}
\newcommand{\redcross}   {\large{\textcolor{red}        {\ding{54}}}}
\newcommand{\blackcross} {\textcolor{black}             {\ding{54}} }

%% file: text/methodology.tex
\section{Experimental Methodology}\label{sec:methodology}

\subsection{Simulation Framework}
\vspace{0.05 in}
\noindent{\bf Simulator.}
\vspace{0.01 in}
We use Scarab~\cite{scarab} for simulation. Scarab uses Intel Pin~\cite{luk2005pin} as the functional model. Main memory is modeled using Ramulator~\cite{kim2015ramulator}. The virtual page size is 4KB with random policy for virtual page to physical frame mapping. Table~\ref{table:configuration} shows the configuration parameters.

%\vspace{0.05 in}
\noindent{\bf Workloads.}
\vspace{0.01 in} 
We use the SimPoints~\cite{sherwood2002automatically} methodology to create 200-million instruction representative regions of the SPEC-2017~\cite{spec17} rate benchmarks and GAP Benchmark Suite (GAPBS)~\cite{beamer2015gap}. %The workloads memory access behavior is shown in Table~\ref{table:workloads}. 
Workloads with LLC MPKI $\geq 10.0$ are considered as memory intensive. We simulate a 4-core system with each SimPoint replicated four times.

\begin{figure*}[ht]
    \centering
    \hspace*{-3mm}
    \includegraphics[width=1\linewidth]{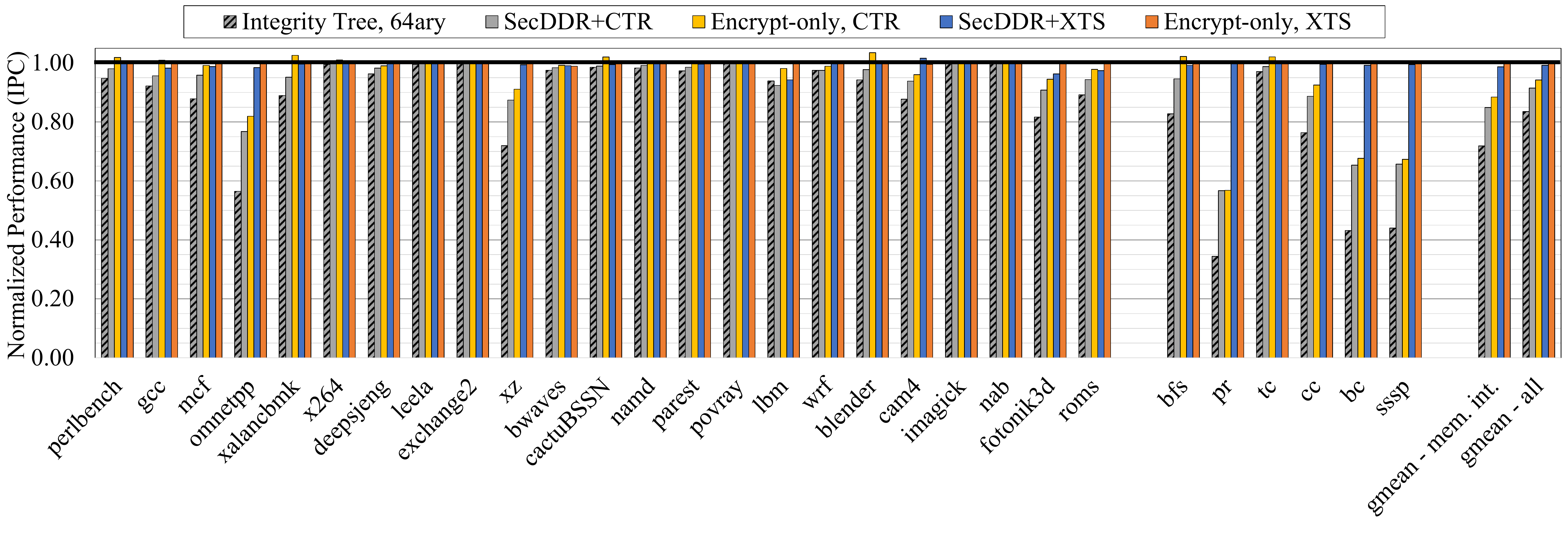}%\vspace*{-3mm}
    \caption{Performance results.\revision{ Normalized performance (IPC) to the Intel TDX baseline.}}
    \label{fig:performance-results}
    \vspace*{-4mm}
\end{figure*}

\begin{table}[ht]
\footnotesize
\centering
\caption{Configuration Parameters}
\label{table:configuration}
\begin{tabularx}{\linewidth}{|c||X|}
 \hline
 Core &  6-wide fetch/retire Out-of-order, 224 entry ROB, 97 entry RS, TAGE-SC-L branch predictor, 3.2GHz, 4 cores \\ 
 \hline
 L1 Cache           & Private 32KB d- \& 32KB i-cache, 64B line, 4-way\\\hline
 Last Level Cache   & Shared 4MB, 64B line, 16-way                      \\\hline
 Prefetcher         & Stream Prefetcher                                 \\\hline
 Metadata Cache     & Shared 128KB, 64B line, 8-way                     \\\hline
 Security Mechanisms & 40 processor-cycles encryption and MAC\\\hline
 Main Memory & 16GB DRAM, 1 channel, 2 ranks, 4 bank-groups, 16 banks, 8Gb\_x8. 64 Read- and 64 Write-entry memory controller queues.\\\hline
 Memory Timings & DDR4-3200 at 1600MHz, tCL\slash tCCDS\slash tCCDL\slash tCWL\slash tWTRS\slash tWTRL\slash tRP\slash tRCD\slash tRAS = 22\slash 4\slash 10\slash 16\slash 4\slash 12\slash 22\slash 22\slash 56 cycles\\
 \hline
\end{tabularx}
%\vspace*{-4mm}
\end{table}

\subsection{Evaluated Systems}

We compare all configurations with a secure baseline that provides memory encryption and integrity protection, however, it lacks replay attack protection, to resemble Intel TDX as the state-of-the-art secure memory design in industrial products. Except in the encrypt-only configurations, MACs are placed in the ECC chips~\cite{inteltdx,fakhrzadehgan2014safeguard}. We consider both counter-mode and AES-XTS encryption modes in our evaluation because they offer a security--performance tradeoff. 

We compare 5 main system configurations. We also compare \papername{} directly to InvisiMem~\cite{aga2017invisimem} in Section~\ref{sec:compare_active_memories}.

\vspace{0.05 in}
\noindent{\bf 1) Baseline:}
\vspace{0.01 in}
The baseline secure system that follows recent academic work with a 64-ary integrity tree and counter-mode encryption. The integrity tree is built on the encryption counters with on-chip caching for both the tree and the encryption counters. We assume an idealized tree and encryption counters with no counter overflow. The encryption-counter lines and the tree nodes have the same number of counters (64). We allow parallel tree-level verification to reduce the overall verification latency. We do not allow speculative use of data. 

\vspace{0.05 in}
\noindent{\bf 2) \papername{}+CTR:}
\vspace{0.01 in} 
\papername{} with the same counter-mode encryption as the baseline.
%We only add a single MAC latency to each memory access.
We assume $\textrm{\textit{OTP}}_{t}$ latency can be hidden. For eWCRC, we increase the write burst length from 8 to 10 ($tBL=5$ for writes). While eWCRC enhances the reliability for all configurations and has only a small performance impact, we make a conservative performance comparison and enable eWCRC \emph{only} for \papername{} configurations. Note that for DDR5 memories the impact of increasing the write burst length is smaller -- from 16 to 18~\cite{micron_ddr5_datasheet}.

\vspace{0.05 in}
\noindent{\bf 3) Encrypt-only, CTR:}
\vspace{0.01 in} 
An upper-bound encryption-only secure system that assumes integrity rather than ensuring it. %We model an \emph{encrypt-only} system without integrity overheads as an upper bound on performance. 

\vspace{0.05 in}
\noindent{\bf 4) \papername{}+XTS:}
\vspace{0.01 in}
A higher-performance variant of \papername{} that avoids counters overhead with AES-XTS encryption.

\vspace{0.05 in}
\noindent{\bf 5) Encrypt-only, XTS:}
\vspace{0.05 in}
An AES-XTS encrypt-only system. %Encrypt-only system with AES-XTS.

To model the AES-XTS encryption overhead, we add the encryption latency to each memory access. AES-XTS does not rely on encryption counters and encryption does not generate any additional memory requests. Although AES-XTS has been adopted as the industry standard for memory encryption~\cite{amdsev, inteltme}, its security guarantees are not identical to those of counter-mode encryption in SGX. Specifically, AES-XTS encryption no longer has  \emph{temporal} variation, meaning if a plain-text line (line-sized block at a specific memory address) has the same value at two different times, it will be encrypted to the same cipher-text, potentially leaking information~\cite{wilke2020sevurity}. Extending \papername{} as described in Section~\ref{sec:proposed_method} to operate with counter-mode encryption is straightforward: Encryption counters are stored and cached as in the baseline secure system but their integrity is protected using per-line MACs, just like data. \ignore{(Figure~\ref{fig:counter-line_protection}).

\begin{figure}[ht]
    \centering
    \includegraphics[width=0.8\linewidth]{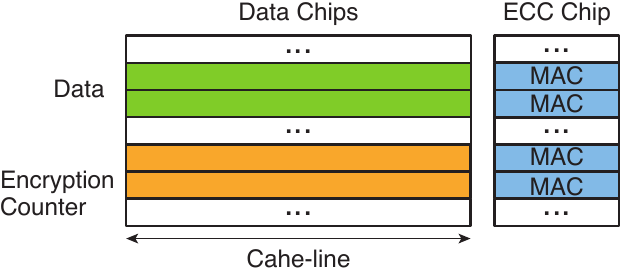}
    \caption{Encryption counter protection in \papername{}.}
    \label{fig:counter-line_protection}
    %\vspace*{-3mm}
\end{figure}
}

%% file: text/evaluation.tex
\section{Evaluation}\label{sec:evaluation}
\subsection{Performance Results}

\vspace{0.05 in}
\noindent{\bf Overall Performance.}
\vspace{0.01 in}
Figure~\ref{fig:performance-results} shows the performance of the different configurations (total IPC) normalized to the TDX baseline. \papername{}+CTR improves average performance by 9.6\% relative to the 64-ary tree baseline, performing within 3.0\% of the encrypt-only system with counter-mode encryption. For memory intensive benchmarks, the average improvement is 18.0\%. The largest speedup is observed for \textit{pr}, \textit{bc}, \textit{sssp}, \textit{omnetpp}, and \textit{xz} that gain 64.7\%, 51.2\%, 49.4\%, 35.9\%, and 21.5\%, respectively. These applications exhibit random memory access patterns and thus low locality. As a result, each data access requires traversing a different branch of the tree, which makes the metadata cache less effective, resulting in multiple off-chip accesses per data access for integrity tree traversal. Only \textit{lbm} exhibits a small slowdown of 1.6\% because it is write-intensive
%(Table~\ref{table:workloads}, Write-PKI)
and is penalized by the longer write burst length of eWCRC that we only add to \papername{}.

\vspace{0.05 in}
\noindent{\bf Encryption Modes.}
\vspace{0.01 in}    
Using AES-XTS, \papername{}+XTS provides 18.8\% average performance improvement relative to the integrity-tree baseline and 5.4\% better than counter-mode encrypt-only because it eliminates the overhead of accessing counters (37.7\% and 11.6\% for memory intensive workloads, respectively). \papername{}+XTS has negligible overhead ($<$1\%) compared with the encrypt-only system (with XTS), which is mainly caused by the extra write burst length due to eWCRC. \textit{cam4} shows slight speedup (1.6\%) with \papername{}+XTS, which appears to be from fewer prefetch requests due to better timeliness. However, this does not change the overall conclusion and we do not consider as an improvement of \papername{}.

\begin{figure}[h]
    \centering
    %\vspace*{-5mm}
    \hspace*{-4mm}
    \includegraphics[width=1.0\linewidth]{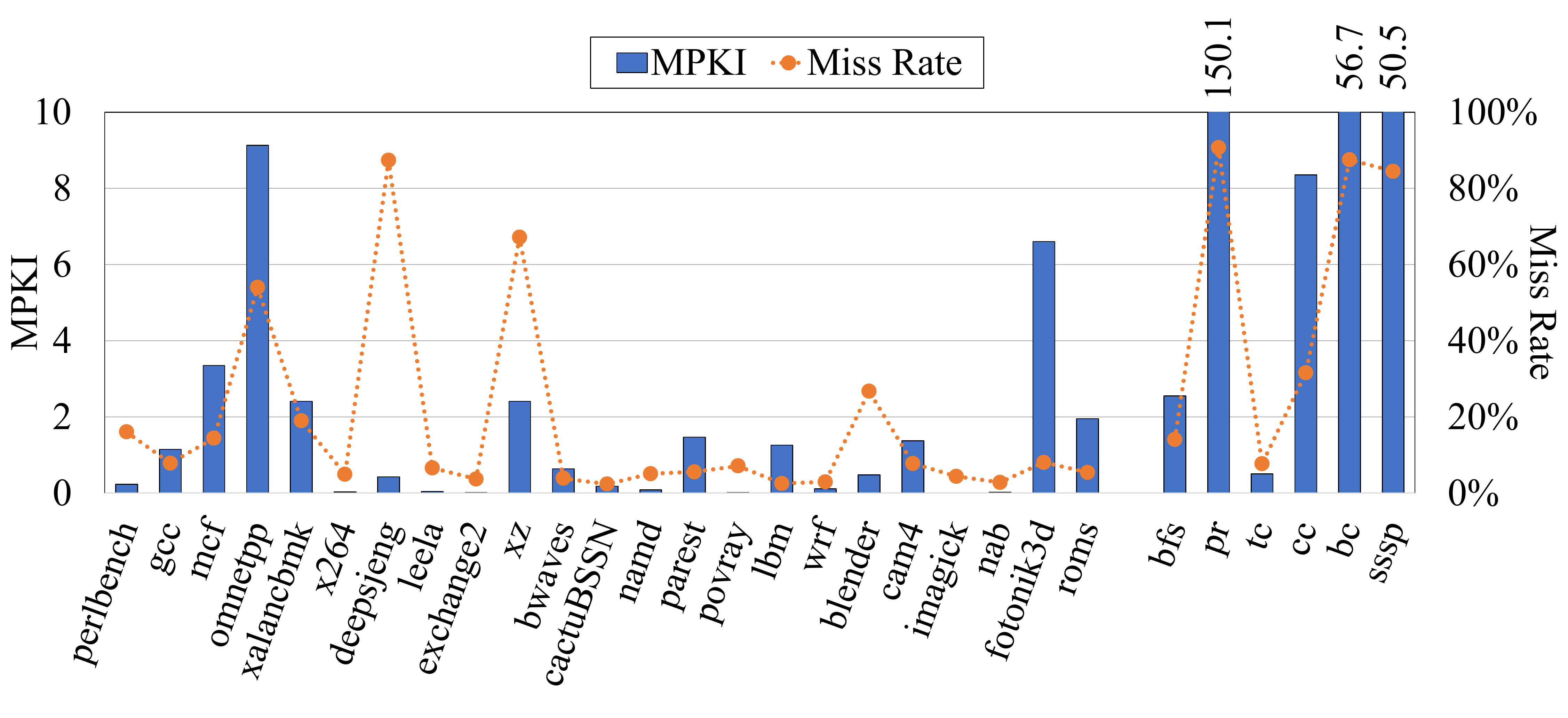}
    \caption{Metadata cache behavior.}
    \label{fig:metacache}
    %\vspace*{-3mm}
\end{figure}

A few benchmarks (\textit{perlbench}, \textit{gcc}, \textit{xalancbmk}, \textit{x264}, \textit{cactuBSSN}, \textit{blender}, \textit{bfs}, and \textit{tc}) exhibit higher performance with counter mode. These benchmarks have high locality and the counter cache has low enough miss rate (Figure~\ref{fig:metacache}) such that the latency saved by pre-computing $\textrm{\textit{OTP}}$ and hiding the encryption/decryption latency outweighs the overhead of fetching counters from memory. In contrast, AES-XTS never accesses counters but always incurs the encryption latency.

Note that in this comparison, the security guarantees of encryption modes in SecDDR+XTS and the integrity tree are not identical. However, assuming AES-XTS is acceptable for the security demands, applying state-of-the-art counter-based trees would be infeasible, and one needs to use hash-based Merkle Trees, which drastically hurts performance. 

\ignore{
\begin{table*}[!htb]
  \footnotesize
  \centering
  \caption{AES engine power overhead.}
  \label{table:storage}
  
  \begin{tabularx}{15cm}{|c||*{2}{>{\centering\arraybackslash}X|}|*{2}{>{\centering\arraybackslash}X|} }
    \hline
    DIMM configurations and                 & \multicolumn{2}{c||}{DDR4-3200, 1600MHz, 1.2V}   & \multicolumn{2}{c|}{DDR5-8800, 4400MHz, 1.1V}\\\cline{2-5}
    AES engine parameters                   &  x4 4Gb   & x8 8Gb                & x4        & x8                    \\ \hline \hline
    AES engines in one ECC chip             &  2        & 3                     & 3         & 3                  \\\hline    
    AES engines power in one ECC chip (mW)  &  70.8     & 106.3                 & 89.3      & 89.3               \\\hline
    %DRAM chip power (mW)                    &  290      & 351.9                 & 252.3     & 306.2              \\\hline
    16GB dual rank power (mW)               &  13230    & 9120                  & 11510     & 7934               \\\hline
    \textbf{Overhead per rank}              &  1.1\%    & 2.3\%                 & 1.6\%     & 2.3\%              \\\hline
  \end{tabularx}
\end{table*}
}

\begin{figure}[ht]
    \centering
    %\vspace*{-4mm}
    %\includegraphics[width=\linewidth]{graphs/arity_sensitivity_dram_gapbs.pdf}
    \includegraphics[width=.95\linewidth]{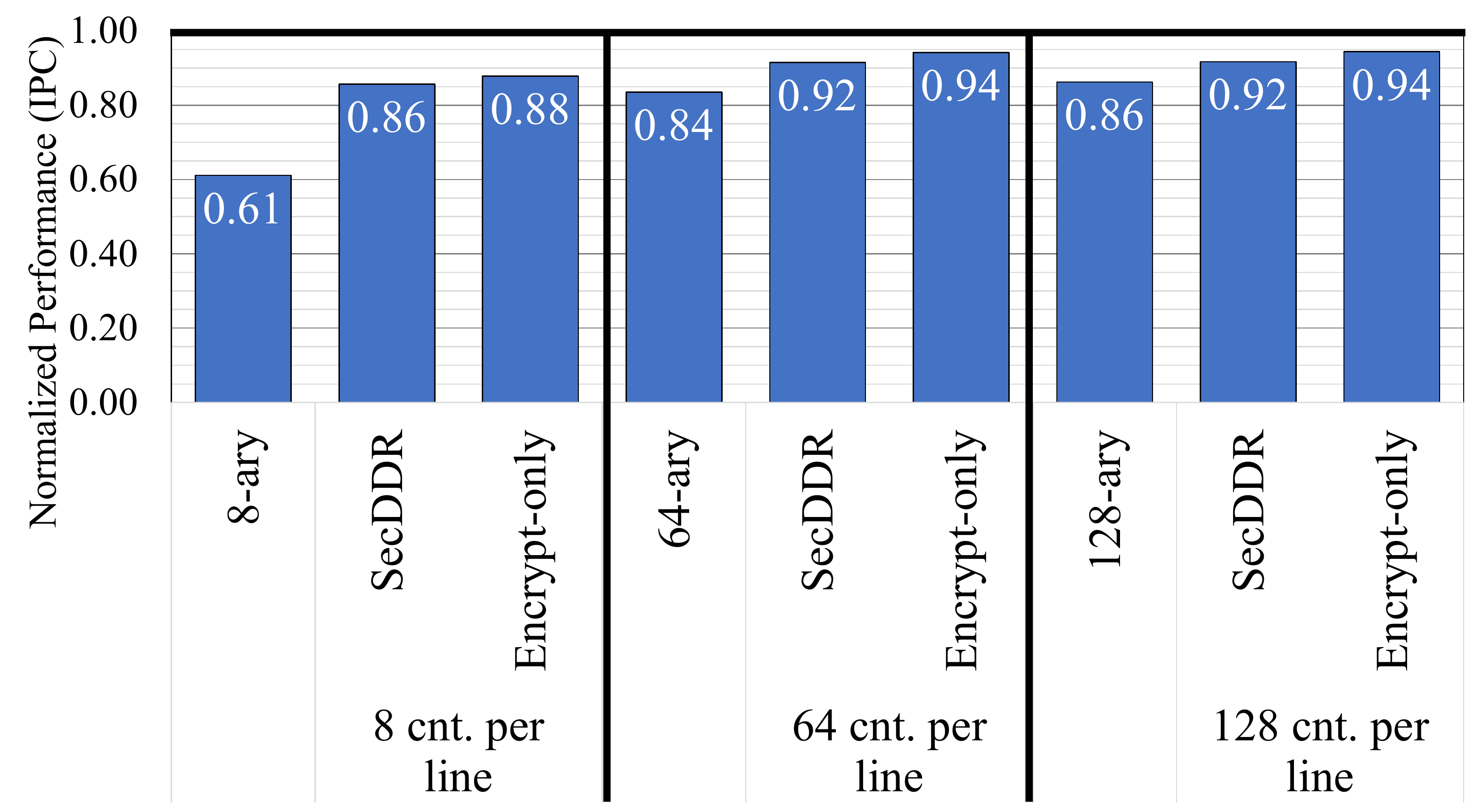}
    \caption{Sensitivity to tree-arity and counter-packing. 
    %\mattan{Does this figure need to change to reflect the new choice of baseline?}
    }
    \label{fig:arity-sensitivity}
    %\vspace*{-2mm}
\end{figure}

\vspace{0.05 in}
\noindent{\bf Sensitivity to Tree Arity.}
\vspace{0.01 in}
Figure~\ref{fig:arity-sensitivity} provides a comparison of different arity values to represent different tree types. The 128-ary design represents the state-of-the-art counter-based tree, MorphTree~\cite{saileshwar2018morphable}, which removes one level of the tree at the cost of greater complexity compared to our 64-arity baseline. Compared with the 128-ary tree, \papername{}+CTR performs 6.3\% better on average and has the advantage of scaling to high-capacity memories. Encrypt-only configurations with 64- and 128-counter exhibit similar performance. With 128-packing, each counter line spans two adjacent 4KB frames, but the random page mapping of our evaluation limits this advantage. 

We also evaluate the 8-ary design to represent hash-based Merkle Trees, which compute the hash over MACs rather than counters and can be used with AES-XTS. However, the performance penalty is very severe, incurring a 38.8\% slowdown. Note that in addition to its compatibility with AES-XTS, the 8-ary design also differs in its placements of MACs. Instead of placing MACs in the ECC chips~\cite{fakhrzadehgan2014safeguard,inteltdx}, the 8 MACs that are hashed together in the tree should be placed in a contiguous block in memory. Otherwise, these MACs must be gathered from their respective locations, increasing the overhead. Thus, it is better to place ECC in the ECC chips in this design. The reliability of this organization is not identical to that of our other configurations, though prior work has established that the reliability is similar~\cite{fakhrzadehgan2014safeguard}.

\subsection{Area \& Power Overheads}

\revision{To estimate the area overhead of implementing \papername{}'s security logic on the DRAM die, we follow the methodology of prior work~\cite{yazdanbakhsh2018dram,gu2020ipim} that report numbers based on older generation of $45nm$ technology to account for the lower density and fewer metal layers in \revision{$\leq22nm$} DRAM process technology.

\vspace{0.05 in}
\noindent{\bf E-MAC Generation.}
\vspace{0.01 in}
On the processor, we need a 16-byte register for the secret key ($K_{t}$), an 8-byte counter ($C_{t}$), and an AES unit. In each ECC chip, we need a 16-byte register for $K_{t}$, an 8-byte counter, and an AES unit.
%Although implementing the security logic in the DB is more cost-effective, it \makeblue{does not protect against malicious modules or on-DIMM vulnerabilities.} 
%\makeblue{To estimate the area and power overhead of implementing \papername{}'s security logic on the DRAM die, we use a similar methodology used by prior work~\cite{yazdanbakhsh2018dram,gu2020ipim} that report numbers based on $45nm$ technology to account for lower density and fewer metal layers in DRAM process technology.
%In the untrusted DIMM threat model, we enhance \papername{}'s security by including the security logic in the ECC chip, which needs to be placed on the DRAM die. We estimate the overhead of the security logic in the ECC chip by reporting numbers based on $45nm$ process technology
The AES engine can be implemented with $0.15mm^2$ area overhead using 45nm technology and provides encryption throughput of $53 \textrm{\textit{Gbps}}$ at $2.1\textrm{\textit{GHz}}$~\cite{intel_45nm_aes}. Table~\ref{table:aes_power} summarizes the power overhead of AES engines in \papername{}. To estimate the power, we scale the power linearly assuming $500\textrm{\textit{MHz}}$ DRAM core frequency and round the number of AES engines to meet the transfer rate of the chip, which is $12.8Gbps$ and $25.6Gbps$ for DDR4-3200 x4 and x8 chips, which results in the total of $70.8mW$ and $106.3mW$ extra power, respectively. SecDDR logic is only implemented in 2 out of 18 (x4) or 1 out of 9 (x8) DDR4 chips in each rank (i.e., the ECC chips).\footnote{ECC is also encrypted to avoid leaking the plaintext MAC.} Considering $290$-$350mW$ for one DRAM chip ($9$-$13W$ for a 16GB dual rank DDR4 DIMM)~\cite{micron_ddr4_calc_power}, the power overhead is less than 3\% per-rank. For DDR5, an x4 DDR5-8800 chip requires encryption throughput of $35.2Gbps$, which results in the total of $89.3mW$ extra power for 3 AES engines operating at 1.1V. Assuming DDR5 memory consumes about 13\% less power than DDR4~\cite{samsung_lower_power}, the total overhead remains below 5\%. Note that this is a conservative estimate given the $10nm$ \emph{class} technology used in DDR5~\cite{kim20191}. 

\begin{table}[!htb]
  \footnotesize
  \centering
  \caption{\revision{AES engine power overhead (powers are in mW).}}
  \label{table:aes_power}
  \begin{tabularx}{.47\textwidth}{|c||*{2}{>{\centering\arraybackslash}X|}}
    \hline
    DIMM configurations and                 & \multicolumn{2}{c|}{DDR4-3200, 1600MHz, 1.2V} \\\cline{2-3}
    AES unit parameters                     &  x4 4Gb       & x8 8Gb\\ \hline \hline
    AES units per ECC chip                  &  2            & 3\\\hline    
    AES power per ECC chip                  &  70.8         & 106.3\\\hline
    DRAM chip power                         &  290          & 351.9\\\hline
    16GB dual rank LRDIMM power               &  13230        & 9120\\\hline
    \textbf{Overhead per rank}              &  2.1\%        & 2.3\%\\\hline
  \end{tabularx}
\end{table}
}

\ignore{
With DDR4 I/O frequency of $1600\textrm{\textit{MHz}}$, SecDDR needs encryption throughput of $23.8Gibps$. Assuming in-DRAM logic can only operate at a limited frequency of $500\textrm{\textit{MHz}}$~\cite{upmem_case_study,pim_samsung_isca21} (i.e., $11.9\textrm{\textit{Gbps}}$ with a 5-cycle AES latency~\cite{intel_45nm_aes}), $2$ AES engines are enough to meet the required throughput. Scaling the power for $500\textrm{\textit{MHz}}$ at $1.2V$, the AES engines require a total of $70.8mW$ extra power. DDR5 memory modules are expected to operate with up to $4400\textrm{\textit{MHz}}$, delivering higher throughput of up to  $8800\textrm{\textit{MTps}}$~\cite{micron_ddr5_datasheet}. At this speed, we need encryption throughput of $65.6Gibps$.}
%\papername{} only uses one 64-bit $\textrm{\textit{OTP}}_t$ every 4 DDR cycles, and thus, we can use one 128-bit AES engine to produce $\textrm{\textit{OTP}}_t$ for two consecutive transactions.}

Thus, we expect the area overhead within the \papername{} device to be $< 1.5{mm}^2$, which is far less than the overhead reported for recent processing-in-memory (PIM) prototypes from memory vendors. These are capable of implementing much more complicated logic on the DRAM die~\cite{upmem_case_study, fim-dimm, skhynix_pim, pim_samsung_isca21}. For instance, recent work~\cite{pim_samsung_isca21} based on $20nm$ DRAM process technology reports $0.712mm^2$ area for a single PIM execution unit, which is more than $20\times$ larger than the AES engine after scaling from $45nm$.

\ignore{
\vspace{0.05 in}
\noindent{\bf Attestation.}
\vspace{0.01 in}
Key exchange and message signing require elliptic curve scalar multiplication and a hash function. Using $45nm$ technology, the multiplier can be implemented with $0.0209mm^2$ area overhead and operate at $1mW$ power~\cite{intel_45nm_galois}. The hash function (e.g., SHA-256) can be implemented with $0.0625mm^2$ area overhead with $1mW$ power overhead~\cite{intel_45nm_sha}.
}

\vspace{0.05 in}
\noindent{\bf Attestation.}
\vspace{0.01 in}
Key exchange and message signing require elliptic curve scalar multiplication and a hash function. Using $45nm$ technology, the multiplier can be implemented with $0.0209mm^2${~\cite{intel_45nm_galois}} and the hash function (e.g., SHA-256) with $0.0625mm^2${~\cite{intel_45nm_sha}} area overhead. \revision{At 1.1V and the peak operating frequency, these units consume $74mW$ and $50mW$ power at $3\textrm{\textit{GHz}}$ and $1.4\textrm{\textit{GHz}}$, respectively. Similar to the AES engine, scaling the power to $500\textrm{\textit{MHz}}$, these units require $14.2$ and $21mW$ extra power, respectively. Note that these units are only needed for attestation during system initialization, and can be turned off otherwise.}

\vspace{0.05 in}
\noindent{\bf Multi-Channel Setting.}
\vspace{0.01 in}
\papername{} works for each memory controller independently, replicating the above logic per channel. We target untrusted DIMMs and each ECC chip in each rank implements the security logic.
%For the untrusted DIMM model, these units need to separately provisioned for each rank, which is a total of $\textrm{\textit{Number of Channels}} \times$ $\textrm{\textit{Number of Ranks per Channel}}$ replications.

%% file: text/compare_active_memory.tex
\section{Comparison with InvisiMem~\cite{aga2017invisimem}}\label{sec:compare_active_memories}
Using mutual authentication to protect the memory bus has been proposed by InvisiMem~\cite{aga2017invisimem} for 3/2.5D-stacked memories. However, adapting InvisiMem to work with commodity DIMMs that operate using the DDRx protocol is impractical and requires fundamental changes to the memory system architecture. Although \papername{} set similar goals as InvisiMem, our approach is effective in addressing these challenges, making \papername{} suitable for adoption. In this section, we first provide a short background on InvisiMem and then, discuss these challenges.

\subsection{InvisiMem}
InvisiMem\footnote{InvisiMem has two flavors. \textit{InvisiMem\_far} is related to this work.}\cite{aga2017invisimem} uses the compute capability in 3/2.5D-stacked memories and the packetized protocol of the Hybrid Memory Cube (HMC) for address/command obfuscation and data confidentiality and integrity protection. InvisiMem extends the TCB to include the HMC logic layer (which contains the memory-side security logic) to create a mutually authenticated channel between the processor and HMC. InvisiMem takes advantage of the packet size flexibility~\cite{hmc_gen2} and adds new metadata to each HMC packet: a payload MAC that is used on the receiving end (processor on reads and memory on writes) to verify packet integrity and freshness.

InvisiMem deems physical attacks on the communication between the logic layer and DRAM cells impractical since in stacked memories, these connections go through the silicon layers (using \revision{through-silicon vias (TSV)} or silicon interposer). This allows InvisiMem to effectively eliminate the integrity tree as the TSVs do not need replay-protection, and thus, storing a MAC with the data is sufficient to protect it while at rest (e.g., against Row-Hammer). On the memory side, after receiving a write request and verifying packet integrity, the security logic within the HMC generates a new MAC and stores it with the data to protect data at rest. On reads, the security logic first reads and verifies the stored MAC and then generates a new (channel) MAC for the packet using the transaction timestamp ($C_t$). Unlike InvisiMem, \papername{} avoids a mutually-authenticated channel and leaves all authentication to the processor, reducing complexity and latency in memory.

\ignore{ , as shown in Figure~\ref{fig:hmc-packet}b and c. 
\begin{figure}[ht]
    \centering
    \hspace*{1mm}
    \includegraphics[width=\linewidth]{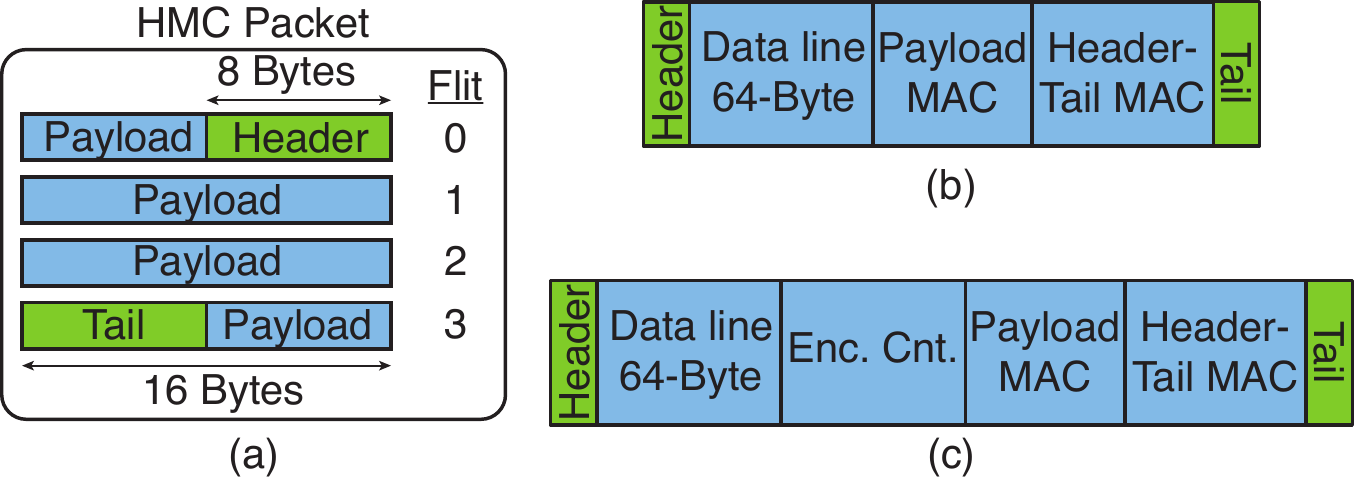}
    \caption{(a) HMC packet~\cite{hmc_gen2} (b) InvisiMem Request Packet (112B) (c) InvisiMem Response Packet (120B).}
    \label{fig:hmc-packet}
\end{figure}
}

\ignore{
\vspace{0.05 in}
\noindent{\bf 2) ObfusMem:}
\vspace{0.01 in}
ObfusMem~\cite{awad2017obfusmem} was developed at the same time as InvisiMem and studies address and command bus obfuscation using a point-to-point mechanism between the processor and the memory. However, data integrity is delegated to integrity trees, which limits the scalability of ObfusMem to large memories.
}

\ignore{
\subsection{Modern DIMM Architectures}\label{sec:buffered-dimm}
Buffered memory modules are designed to provide high-speed large-capacity memories. The vast number of memory devices on a module increases the capacitive load on the memory bus, and the module interconnects, which adversely impacts signal stability and integrity. To mitigate this problem, the industry has adopted buffered memory modules, which decouple the I/O signals to each of the DRAM devices by adding a buffer chip to the module. \textit{Registered DIMMs (RDIMM)} and \textit{Load-Reduced DIMMs (LRDIMM)} are the two types of buffered DIMMs. While RDIMMs only buffer the Command, Control, Clock, and Address (CCCA) signals, LRDIMMs buffer both CCCA and data signals. 
}

\subsection{Challenges of Adapting InvisiMem for DDRx DIMMs}\label{sec:bad_invisimem}

Adapting InvisiMem for commodity DDRx DIMMs introduces challenges that are not trivial to address.

\vspace{0.05 in}
\noindent{\bf Trusted HMC vs. Untrusted DIMM.}
\vspace{0.01 in}
The security-guarantees in InvisiMem are provided given that the logic layer in the HMC is trusted as part of the TCB and the 3D-stacked DRAM cannot be penetrated. This assumption, however, does not hold for commodity \revision{DDRx} DIMMs. One could adapt InvisiMem for DDR DIMMs by placing the security logic in a discrete component on the DIMM (e.g., a buffer chip). However, this implementation does not protect the DIMM interconnects and other components on the DIMM, leaving the system vulnerable to malicious DIMMs and on-DIMM replay attacks. Thus, in addition to the security logic, the threat model must consider the \textbf{entire} DIMM as trusted (i.e., \emph{trusted DIMM}). %(similar to trusted DIMM model in Section~\ref{sec:tcb}). 

To mitigate this problem, the security logic could be placed in the DRAM \revision{chips}, however, this is infeasible because InvisiMem relies on memory-side integrity verification of every memory transaction; an operation that requires the entire packet payload (i.e., 64-Byte line) to be available. As opposed to an HMC in which all DRAM vaults are connected to a centralized logic layer, in DIMMs, data is distributed across multiple DRAM \revision{chips}, which makes this approach impractical. Alternatively, the processor could create a separate secure channel to each DRAM chip, but this increases cost and requires increasing the burst length to append $\textrm{\textit{MAC}}_t$ to each transaction (on both reads and writes).\footnote{We can also protect the DIMM by implementing an integrity tree on the DIMM and placing its root in the buffer chip, however, this is not a scalable solution.} 

\ignore{
We propose to use SecDDR's insight and only protect the MAC (that is transferred on the ECC bus) by creating a mutually authenticated channel \emph{only} to the ECC chips. While this reduces cost by limiting modifications to the DRAM devices, longer bursts are still required to transfer $\textrm{\textit{MAC}}_t$ of the stored MAC and additional latency is necessary for memory-side authentication (incurring up to 14\% slowdown relative to \papername{} as we discuss in Section~\ref{sec:ivisimem-perf-comp}).\footnote{We can also protect the DIMM by implementing an integrity tree on the DIMM and placing its root in the buffer chip, however, this is not a scalable solution.}
}

As a result, the approach proposed by InvisiMem is not suitable for commodity DDRx DIMMs that can contain malicious components and can be easily tampered with. On the other hand, even if the threat model with a trusted DIMM is an acceptable option, applying InvisiMem still requires fundamental changes in the design constraints of a modern \revision{DDRx} DIMM, which we discuss next.

\vspace{0.05 in}
\noindent{\bf Packetized Protocol vs. DDRx.}
\vspace{0.01 in}
InvisiMem is particularly designed based on the packetized protocol in HMC, which has been deprecated~\cite{hmc_no_support}. However, DDR is not packetized, and given how widely it is accepted as an industry standard, changing this protocol is extremely difficult.

\vspace{0.05 in}
\noindent{\bf Additional Latency \& Changes to the DDR Timing Parameters.}
\vspace{0.01 in}
InvisiMem's memory-side integrity verification adds additional latency to each memory access and requires changing the DDR timing parameters. Specifically,  \texttt{tCL} must grow to account for the MAC latency. While this additional latency is deterministic (i.e., the MAC latency), it is on the critical path of every memory access. 

\vspace{0.05 in}
\noindent{\bf Reduced Data-Rate due to the Centralized Buffer Chip.}
\vspace{0.01 in}
Implementing the memory-side verification and delegating the integrity-protection to the memory requires introducing a \textit{centralized} data buffer on the memory module. That is, for memory-side MAC generation and verification, the memory module has to gather {\em all} 64 bytes of a line to compute its MAC. The centralized buffer, however, lowers the achievable memory frequency and bandwidth~\cite{idt_lrdimm, next_platform_lrdimm}.

\begin{figure}[th]
    \centering
    %\vspace*{-5mm}
    \includegraphics[width=.9\linewidth]{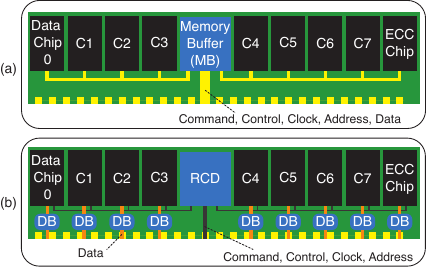}
    \caption{Comparison between old (a) and new (b) DIMM architectures. (a) Centralized Memory Buffer in DDR3. (b) Centralized buffer for Control Signals and distributed Data Buffers in DDR4 and DDR5.}
    \label{fig:buffered_dimm}
    %\vspace*{3mm}
\end{figure}

\begin{figure*}[t]
    \centering
    \hspace*{-3mm}
    %\vspace*{-3mm}
    %\includegraphics[width=\linewidth]{graphs/compare_invisimem_gapbs_memint.pdf}
    \includegraphics[width=\linewidth]{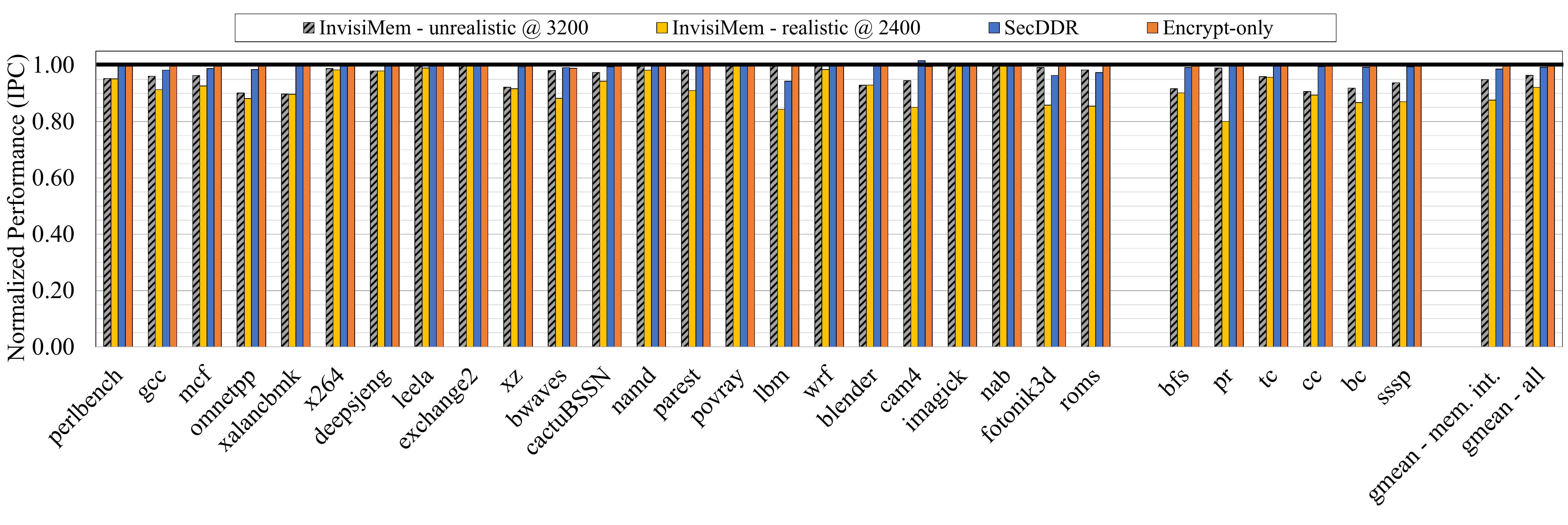}%\vspace*{-3mm}
    \caption{Performance comparison with InvisiMem using realistic and unrealistic memory frequencies. \revision{Normalized performance (IPC) to the Intel TDX baseline. All configurations use AES-XTS encryption.}}
    \label{fig:compare_invisimem}
\end{figure*}

Figure~\ref{fig:buffered_dimm}a shows a DDR3 LRDIMM architecture with a centralized memory buffer chip (MB). All CCCA and data signals are first routed to the MB and then routed from the MB to each DRAM \revision{chip}~\cite{idt_lrdimm, chameleon}. The distance disparity (and thus, the latency) between different DRAM \revision{chips} to the centralized MB limits the highest data rate and the frequency at which the \revision{DIMM} can operate~\cite{idt_lrdimm, next_platform_lrdimm, chameleon}. To address this problem, the buffer chips for the CCCA and data in DDR4 and DDR5 LRDIMMs are distributed~\cite{idt_lrdimm, rambus_ddr5_db}, as shown in Figure~\ref{fig:buffered_dimm}b. As discussed in Section~\ref{sec:threat-model}, The CCCA signals are buffered in the RCD chip, whereas the data is buffered in \emph{distributed} DB chips. Distributed buffers are at a short and identical distance from their corresponding DRAM \revision{chips} across the module, which reduces the buffering latency, enabling higher data rates. Adding a centralized buffer is undesirable and contradicts the main reason that newer memory technologies have transitioned to distributed data buffers.

\ignore{
\begin{tcolorbox}
{\bf Key Takeaway:} Existing mutual authentication proposals rely on centralized data authentication and verification on the memory module. Centralized data buffers are unfit for DDRx security for two reasons: (1) this approach is impractical as it leaves memory modules vulnerable to on-DIMM attacks, and (2) this limits the operating frequency of the module and degrades system performance.
\end{tcolorbox}
}

\subsection{\papername{}'s Compatibility with Trusted Memory Modules}\label{sec:secddr_with_trusted}
To provide an iso-secure baseline that mimics the case in which InvisiMem's~\cite{aga2017invisimem} security logic is placed in a discrete component on the DIMM, we consider \papername{} with a \emph{trusted} module that assumes on-DIMM attacks to be impossible, and discuss the placement of the on-DIMM security logic and the TCB components accordingly.

\begin{figure}[h]
    \centering
    %\vspace*{-3mm}
    %\includegraphics[width=\linewidth]{graphs/tcb.pdf}
    \includegraphics[width=.95\linewidth]{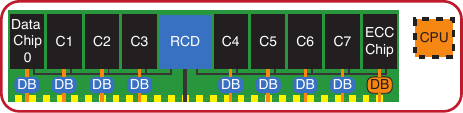}
    \caption{\papername{}'s TCB with a trusted DIMM. CPU and the DIMM are in the TCB. ECC DB contains the security logic and acts as the DIMM's root-of-trust.}
    \label{fig:tcb_trusted_memory}
    %\vspace*{-3mm}
\end{figure}

As shown in Figure~\ref{fig:tcb_trusted_memory}, we can apply \papername{} to a trusted module by placing the security logic inside the data buffer (DB) of the ECC chip(s). The ECC DB acts as the DIMM's root-of-trust, and is responsible for attestation and establishing the secure E-MAC channel. We assume the entire DIMM is in the TCB. Note that an attacker can perform a man-in-the-middle replay attack by tampering with the DIMM interconnects or using a malicious DIMM that contains a hardware \emph{trojan}. These attacks are impractical in the 3D-stacked HMC context of InvisiMem, but HMCs have proven non-viable in the market.% as the TSVs are orders of magnitude smaller than the DIMM interconnects~\cite{aga2017invisimem}.}

\subsection{Performance Comparison with InvisiMem}\label{sec:ivisimem-perf-comp}
\papername{}'s E-MACs address all the above challenges. In contrast with memory-side MAC generation in InvisiMem, E-MACs are computed with no significant latency on the memory access critical path as the $\textrm{\textit{OTP}}_t$ can be pre-computed ahead of time. Furthermore, to encrypt MACs, counter-mode encryption only requires XOR-ing the MAC with the $\textrm{\textit{OTP}}_t$ that is performed cycle-by-cycle in the ECC DB or the ECC chip independently, and does not require any data communication or synchronization with the data chips. Thus, \papername{} does not require any centralized buffering, and it is compatible with the existing LRDIMMs that use distributed DBs.

\begin{figure*}[t]
    \centering
    \hspace*{-3mm}
    \includegraphics[width=\linewidth]{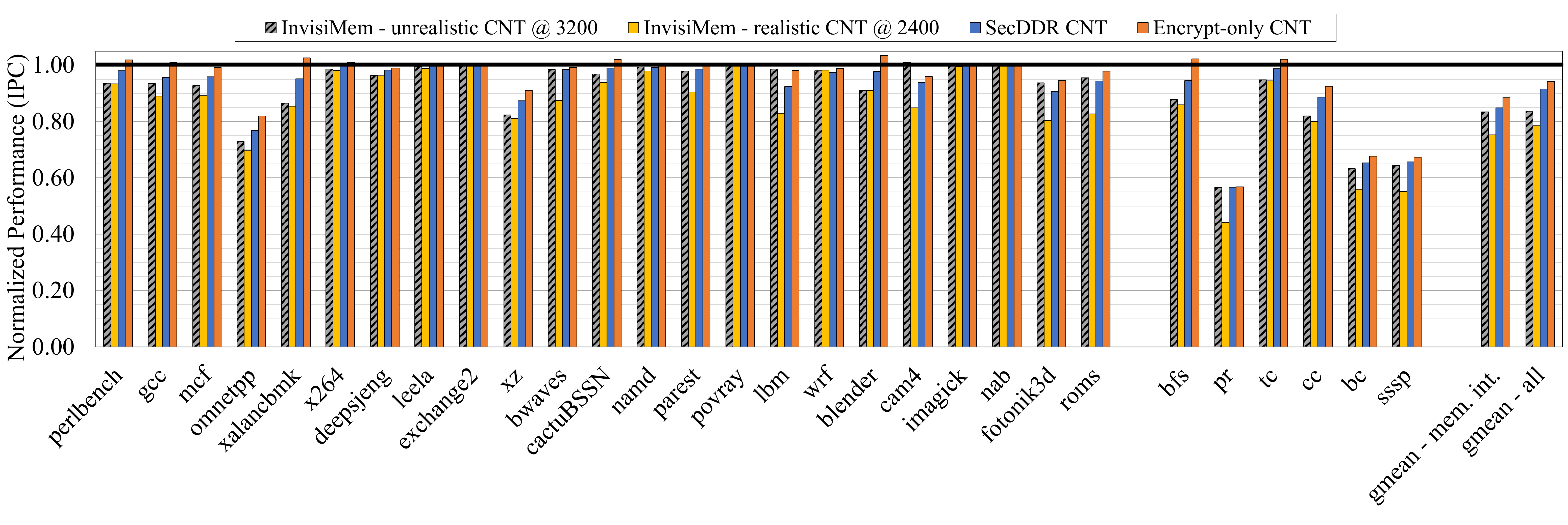}
    \caption{\revision{Performance comparison with InvisiMem using realistic and unrealistic memory frequencies. Normalized performance (IPC) to the Intel TDX baseline. All configurations use counter-mode encryption with 64 counters per-line.}}
    \label{fig:compare_invisimem_counter}
    %\vspace*{-3mm}
\end{figure*}

Figure~\ref{fig:compare_invisimem} shows the performance comparison between \papername{} and InvisiMem. %\makeblue{For a fair comparison, we consider an iso-secure threat model assuming a trusted DIMM by placing the security logic of both designs in one of the buffer chips (MB for InvisiMem and DB of the ECC chip for \papername{}). Note that moving \papername{}'s security logic from the ECC chip(s) to the ECC DB does not affect performance as the $\textrm{\textit{OTP}}_t$ latency can be hidden}. \mattan{This is repeated from before, why not just cut the fair baseline bit from here?} 
We assume AES-XTS in all configurations. Only \papername{} is equipped with eWCRC and incurs the longer bursts necessary for it. To evaluate InvisiMem (with a \emph{trusted} DIMM), we consider two cases: \emph{unrealistic} and \emph{realistic}. In the unrealistic case, we assume the memory can operate at 1600MHz (3200MT/s) and InvisiMem's overhead is only due to the $2\times$ MAC latency on the access critical path (one on the processor and one on the DIMM). Although this configuration has a small average overhead of 2.9\% relative to \papername{} (3.8\% on memory intensive applications), it is not achievable because the memory must run at a lower frequency due to the centralized data buffer. The realistic implementation operates at 1200MHz (2400MT/s) to account for this and InvisiMem then incurs a 7.2\% average performance overhead relative to \papername{} (11.2\% for the memory intensive applications). 
\ignore{
\makeblue{Using our extension of InvisiMem that can operate with \emph{untrusted} modules (Section~\ref{sec:bad_invisimem}), we only mutually-authenticate the MAC transfers to the ECC chips. This requires extending the burst length (as with \papername{}'s eWCRC) on both read and write operations, but also still incurs the $2\times$ MAC latency penalty. This untrusted-DIMM extension we devise increases the slowdown of InvisiMem to 6.2\% on average compared to an untrusted DIMM with \papername{} -- up to 14\% in \textit{pr}.} %\makeblue{In either case, placing InvisiMem's security logic in the buffer chip cannot protect against on-DIMM attacks.}
}

Compared to the unrealistic idealized InvisiMem, \papername{} performs worse on \textit{lbm}, \textit{fotonik3d}, and \textit{roms} by 6.6\%, 3.0\%, and 1\%, respectively. This difference is due to the extra write burst length in \papername{}. %\papername{} does outperform the untrusted-DIMM extension to InvisiMem that also has a longer burst.

\revision{Figure~\ref{fig:compare_invisimem_counter} provides a similar comparison using counter-mode encryption. We observe a similar trend using counter-mode encryption and \papername{} outperforms InvisiMem unrealistic and realistic by 9.4\% and 16.6\%, respectively. Note, however, that AES-XTS is faster and provides higher overall performance (similar to Figure~\ref{fig:performance-results}).}

%% file: text/related_work.tex
\section{Other Related Work}\label{sec:relate-work}\label{sec:state-of-the-art-secure-nvm}
\vspace{0.05 in}
\noindent{\bf Other Active Memory Designs.}
\vspace{0.01 in}
ObfusMem~\cite{awad2017obfusmem} was developed concurrently with InvisiMem and obfuscates the address and command buses with a point-to-point mechanism between the processor and the memory. Data integrity is delegated to integrity trees, limiting its scalability. SecureDIMM~\cite{shafiee2018secure} provides address and command confidentiality by encrypting the memory bus, however, integrity protection is not provided. SecureDIMM uses Freecursive ORAM~\cite{fletcher2015freecursive} to provide on-DIMM security. A similar proposal by Gundu et al. ~\cite{gundu2014case} off-loads integrity trees to the memory. Unfortunately, similar to InvisiMem, these designs rely on centralized buffer chips, which is not applicable to modern modules.

\vspace{0.05 in}
\noindent{\bf Reducing Overheads for Secure Memory.}
\vspace{0.01 in}
Synergy~\cite{saileshwar2018synergy} is a reliability-security co-design that uses ECC to eliminate the MAC bandwidth overhead in ECC-DIMMs. SafeGuard~\cite{fakhrzadehgan2014safeguard} eliminates the storage and memory overhead of parities in Synergy. VAULT~\cite{taassori2018vault} builds a high arity tree by extending split-counters~\cite{yan2006improving} to higher levels of a counter-based tree. Morphable-Counters~\cite{saileshwar2018morphable} is a 128-ary tree that dynamically allocates more bits for frequently updated counters to reduce counter overflow. Taassori et al. ~\cite{taassori2020compact} propose a compact tree design to reduce the parity update overheads in Synergy.

%% file: text/conclusion.tex
\section{Conclusion \& Future Work}\label{sec:conclusion}
Integrity trees are not scalable to protect large-scale memories due to the severe performance overhead. While mutual authentication is promising, existing proposals require fundamental changes in the memory system architecture. In this paper, we show that by identifying practical types of replay attacks, we can provide a low-cost scheme to protect the DDR interface against them. We propose \papername{}, which creates a reply-protected bus by encrypting the MACs. \papername{} has a negligible performance overhead and does not change the underlying DDR protocol. \papername{} can be extended to use the on-DIMM encryption units to encrypt the address and command for traffic obliviousness.